%% file: Graphene_Dirac_analogue_gravity.tex
\documentclass[11.5pt,letterpaper]{article}
\pdfoutput=1 

\input{Structure.tex} 

\title{%
       \vspace{-1.0cm}
       \centering\boldmath\LARGE\bfseries%
       Graphene, Dirac equation and analogue gravity
       \bigskip
       }

\medskip

\author{\small\textsc{Antonio Gallerati}%
\vspace{0.15em}%
}
\affil{%
\makebox[\textwidth][c]{Politecnico di Torino, Dipartimento di Scienza Applicata e Tecnologia, corso Duca degli Abruzzi 24, 10129 Torino, Italy}
}
\affil{Istituto Nazionale di Fisica Nucleare, Sezione di Torino, via Pietro Giuria 1, 10125 Torino, Italy%
}
\affil{\href{mailto:antonio.gallerati@polito.it}{\texttt{antonio.gallerati@polito.it}}%
}

\date{}

\makeatletter
\patchcmd{\@maketitle}{\begin{center}}{\begin{adjustwidth}{-0.5in}{-0.5in}\begin{center}}{}{}
\patchcmd{\@maketitle}{\end{center}}{\end{center}\end{adjustwidth}}{}{}
\makeatother

\begin{document}

\maketitle


\begin{abstract}
{\noindent%
We provide an updated study of some electronic properties of graphene nanoscrolls, exploiting a related curved space Dirac equation for the charge carriers. To this end, we consider an explicit parametrization in cylindrical coordinates, together with analytical solutions for the pseudoparticle modes living on the two--dimensional background.
These results are then used to obtain a compact expression for the sample optical conductivity, deriving from a Kubo formula adapted to the 1+2 dimensional curved space. The latter formulation is then adopted to perform some simulations for a cylindrical nanoscroll geometry.
}%
\end{abstract}


\bigskip

\tableofcontents

\pagebreak



\section{Introduction}\label{sec:Intro}
The discovery of graphene--like materials has determined a great interest deriving form the synthesis of real, prototypical two--dimensional systems \cite{novoselov2004electric,novoselov2005twodimato}.
Graphene, in particular, has emerged as an exciting framework with notable physical properties
\cite{Novoselov:2005kj,stankovich2006graphene,geim2007rise,geim2009graphene}.\par
A graphene layer consists in a single--atom--thick sheet of hexagonally arrayed carbon atoms. It has attracted attention because of its electronic, mechanical and optical characteristics \cite{CastroNeto:2009zz,katsnelson2007graphene,Vozmediano:2010zz}
together with a variety of quantum phenomena that are characteristic of 2D Dirac fermions, such as specific integer and fractional quantum Hall effects or peculiar oscillations exhibiting non--zero Berry's phase \cite{zhang2005experimental,gusynin2005unconventional,novoselov2007room,
bolotin2009observation,tHoke2006fractional,du2009fractional,carmier2008berry}.
All these properties, combined with its stability, flexibility and near-ballistic transport at room temperature, make graphene an ideal material for many nanoscale applications \cite{li2008chemically,stankovich2006graphene,bonaccorso2010graphene,
sun2010graphene,lui2010ultrafast,Andrei:2012my}.
\par
The substrate electronic properties are naturally described in terms of charge carriers that, near the Fermi energy, mimic particles with zero mass and an effective limiting speed. 
It is then possible to observe a relativistic--like particle motion at sub-luminal velocities \cite{Novoselov:2005kj,Gusynin:2006ym,CastroNeto:2009zz}.
We then find the equivalent of a 2D gas of charged particles described by a  relativistic Dirac equation, rather than a non-relativistic Schrödinger equation with an effective mass.
The particle zero--mass description is a robust feature of graphene, being protected, by a certain extent, from the presence of time--reversal and parity symmetry of the system. The introduction of electron-electron interactions or substrate distortions turn out to be quite irrelevant at low energies, and do not alter the massless formulation \cite{roldan2008effect,foster2008graphene,pereira2009tight,Kotov:2010yh}.\par
From the perspective of high energy physics, the existence of observable pseudoparticles moving in a real, curved background is of great interest. The first remarkable consequence is the possibility to examine quantum fields living in a laboratory system which acts as a lower--dimensional curved spacetime \cite{Cortijo:2006xs,katsnelson2007graphene,geim2007rise,Vozmediano:2008zz}. This provides a bridge between condensed matter and theoretical quantum models, where the analogues of many high energy physics effects can be explored in a solid state system \cite{Zhang:2005zz,Gusynin:2006ym,CastroNeto:2009zz,Boada:2010sh,
Vozmediano:2010zz,Iorio:2011yz,Iorio:2012xg,Iorio:2013ifa,Gallerati:2021rtp}.
Moreover, the massless pseudoparticle motion realizes the analog of a relativistic system, with characteristic limiting velocity given by the Fermi velocity  $\vF$ rather than the speed of light $c$.%
\footnote{%
For graphene, one has $\vF\sim c/300$\:.
}
Then, one is naturally led to consider this special relativistic--like behaviour combined with a supporting curved geometric background, the latter obtained by geometrically deforming the graphene layer. This would give rise to a tabletop general relativistic--like system, the charge carriers acting as Dirac pseudoparticles in a lower--dimensional curved spacetime \cite{Birrell:1982ix,Brill:1966tia,Wald:1984rg,Cortijo:2006mh,Cortijo:2006xs,deJuan2007charge,
Vozmediano:2010zz,deJuan:2009ldt,deJuan:2012hxm,Amorim:2015bga,Gallerati:2021rtp,Gallerati:2018dgm,
Gorbar:2007kd,Boada:2010sh,fillion2021numerical,deoliveira2022connecting}. This unique combination therefore offers the possibility of observing quantum fields in a real, curved background, giving rise to many peculiar effects affecting the electronic properties of the substrate. The study of 2D Dirac materials could then provide new insights in theoretical as well as in experimental physics.\par
Another fascinating opportunity would be to exploit the general relativistic nature of a geometrically deformed layer to test some quantum gravity scenarios \cite{novello2002artificial,Barcelo:2005fc,Barcelo:2000tg}.
This approach is dubbed \emph{bottom--up} formulation, and considers suitable condensed matter system as supporting frameworks in which to observe analogues of gravitational effects. The \emph{sine qua non} of any `analogue model' is, in particular, the existence of some emerging effective metric that captures the notion of curved spacetime arising in general relativity \cite{Barcelo:2005fc,Barcelo:2000tg}: the latter effective metric controls the dynamics of quantum particles, and it is then possible to exploit some convenient techniques usually belonging to general relativity (or quantum gravity). On the other hand, there is the possibility that direct observation of unexpected or unconventional effects could lead to suitable variants of these analogue models, which would in turn provide some hints towards new (or more complete) formulations of quantum gravity theories in high energy physics.\par\smallskip
In the following, we will take advantage of the described analogue--gravity approach to investigate some physical features of specific curved configurations for a graphene--like substrate, using the related effective metric to study the pseudoparticle motion.%
\footnote{With `graphene--like materials' we mean 2D materials featuring a honeycomb lattice and an emergent behaviour as Dirac fermions for the pseudoparticles wavefunction \cite{Wang2015rare,Zhu2007simulation,Malko2012twodim}.
}
In particular, we will focus on graphene nanoscrolls optical properties, exploiting a curved space formulation that, in turn, gives rise to a Dirac equation in cylindrical geometry. The latter can be used to describe the dynamics of the fermionic charge carriers, and will be instrumental in determining a simple, explicit form of an adapted Kubo formula for the sample optical conductivity.
%
\bigskip

\section{Dirac formalism}\label{sec:Dirac}
The hypothesis that carbon--based materials may realize the physics of massless Dirac fields dates back to the 1980s \cite{Semenoff:1984dq}, a property that discriminates graphene from other bidimensional systems and makes it a privileged framework for theoretical investigations and experimental applications.\par
The origin of the Dirac formulation in graphene is different from the one in high energy physics.%
\footnote{%
In high energy physics, the Dirac equation emerges from the request of spacetime Lorentz invariance and other general considerations related to special relativity and quantum mechanics.
}
The spinorial nature of the fields in graphene--like materials emerges as a consequence of the lattice topological structure. In the honeycomb lattice, a unit cell is composed of two adjacent atoms belonging to the two inequivalent, interpenetrating triangular sublattices giving rise to the global hexagonal geometry. The same structure is present in the reciprocal momentum space and reflects in the single--electron wavefunction. The latter is conveniently expressed as a two--component spinor, satisfying a 2D Dirac Hamiltonian describing massless fermions \cite{Novoselov:2005kj,Gusynin:2006ym,geim2009graphene}, the characterization being not affected by lattice deformations retaining the topological configuration \cite{CastroNeto:2009zz,Kotov:2010yh}.\par

\paragraph{Electronic structure.}
The graphene honeycomb crystal structure is characterized by two types of C–C bonds (denoted by $\sigma$, $\pi$) constructed from the four valence orbitals. In particular, three in--plane $\sigma$ bonds join every C atom to its three neighbours. An additional type of bond is associated with the overlap of out--of--plane $\pi$ orbitals (or $p_z$ orbitals, $z$ being the direction perpendicular to the sheet). In general, one expects the electronic wavefunctions from different lattice atoms to overlap. However, there is no overlap between the $p_z$ orbitals and the $s$, $p_x$, $p_y$ orbitals involved in the $\sigma$ bonds. Consequently, the $p_z$ electrons, which form the $\pi$ bands in graphene, can be treated independently from the other valence electrons \cite{charlier2007electron}.\par
The pseudorelativistic characterization of the $p_z$ charge carriers is manifest in the reciprocal momentum space, where the energy turns out to be a linear function of the quasi-momentum at the vertices of the first Brillouin zone (FBZ). The latter, in its turn, has an hexagonal structure, rotated by a $\pi/2$ angle with respect to the honeycomb lattice. The six corners of the FBZ hexagons are subdivided into two inequivalent classes, since only two are truly independent, while the other four can be reached by means of a reciprocal lattice vector. The mentioned vertices of the hexagonal FBZ are referred to as \emph{Dirac points}.\par
The substrate intrinsic Fermi surface is reduced to the six Dirac points of the FBZ, where the $\pi$ band electronic dispersion is found to be linear, like in relativistic massless particles. This is in contrast with the usual, quadratic energy–momentum relation obeyed by electrons at band edges in conventional semiconductors. Graphene turns then out to be a special semimetal (zero--gap semiconductor), featuring massless Dirac charge carriers with relativistic--like behaviour with respect to $\vF$\,. In particular, three--dimensional plots of the two--dimensional linear bands produce Dirac cones, whose conical apices touch at the FBZ corners.
The Fermi level is then taken as the zero--energy reference, and the Fermi surface is defined by means of the Dirac points. Since only the bonding $\pi$ states lie in the vicinity of the Fermi level, the $\sigma$ bonds are usually neglected in the calculations of the electronic properties of graphene around the Fermi energy.%
\footnote{%
The $\sigma$ bands are well separated in energy (more than 10 eV in the center of the FBZ) and are neglected in calculations, since they are too far away from the Fermi level to play a role.
}
The electronic properties are then determined by the $\pi$ orbitals, which give rise to wide electronic valence and conduction bands.
Within the discussed approximation, it is then easy to describe the electronic spectrum of single--layer graphene in terms of a rather simple tight--binding Hamiltonian, leading to analytical solutions for their eigenstates \cite{charlier2007electronic,Gallerati:2018dgm}.%
\footnote{%
A tight--binding model describes electrons in the limit of far apart ions, the single--particle eigenstates referring to a charge carrier affected by a single ion; this results in a set of lattice sites with a single--level state, where the electrons can tunnel to their first neighbour atoms only.
}
\par
When a graphene sheet is bent, the bond angles between the $\sigma$ and $\pi$ orbitals varies, introducing peculiar curvature effects into the electronic properties of the material. We will analyse this occurrence more in detail in the following sections.

\paragraph{Continuum limit.}
Graphene can be warped in the out--of--plane direction, similar to a sheet of paper \cite{xie2009controlled,kim2011multiply}.
The electronic structure of graphene is directly affected by curvature \cite{fogler2010effect}, the specific folding inducing various and intriguing physical effects.
The analysis of certain curved configurations can bring out some peculiar features of the pseudoparticles, originating from the discussed massless behaviour in the curved substrate \cite{osipov2005electronic,kolesnikov2006continuum,
morpurgo2006intervalley,lee2009surface}. A suitable choice of the sample geometry and related parametrization would give rise to some remarkable (and detectable) physical properties.\par
The quantum Dirac formulation is fully realized for the quasiparticles when restricting to low--energy excitations: for energy regimes well below $E_\ell\,\sim\,\vF/\ell$, the charge carriers wavelength is large with respect to the lattice average parameter $\ell\sim0.142\,\mathrm{nm}$, so that the pseudoparticles do not perceive the discrete layer nature, then justifying the continuum picture.\par
Furthermore, electrons featuring very large wavelengths are more susceptible to substrate (spacetime) curvature, again suggesting to limit ourselves to low--energy ranges, where the quasiparticles wavelengths are large if compared with the honeycomb dimensions,
\begin{equation}
\lambda\:>\:2\pi\,\frac{\vF}{E_\ell}\:\sim\:2\pi\ell\:.
\end{equation}
Finally, we should also demand that the curvature be small with respect to a limiting value, related to the lattice dimension: otherwise, the geometric formulation cannot be given in terms of a smooth metric.
The peculiar lattice structure is then naturally combined with the possibility to move through suitable energy regimes, where some discrete aspects and the continuum spacetime picture coexist.
%
%
\medskip

\subsection{Curved space Dirac equation}
Graphene--like \emph{flat} layers are two--dimensional analogues of relativistic systems, the role of limiting speed being played by the Fermi velocity $\vF$\,. The dynamics of the $\psi$ pseudoparticles in the long wavelength regime (continuum limit) is given in terms of a massless Dirac action that, dropping spinorial indices, we write as
\begin{equation}
\mathcal{S}_0\=i\,\hbar\,\vF \int{d^{3}x\;\bar{\psi}\;\gammafl^a\,\dd[a]\psi}\;,\qqquad
\label{eq:flatDiracaction}
\end{equation}
where $a=1,2,3$ is the flat spacetime index.%
\par
A suitable set of $\gammafl$ matrices for a 1+2 flat dimensional background can be given in terms of the Pauli matrices as
\begin{equation}
\gammafl_a\=\big(\,\sigma_3\,,\;-i\,\sigma_1\,,\;i\,\sigma_2\,\big)\;.
\label{eq:gammaflat}
\end{equation}
The $\gammafl^a=\eta^{ab}\,\gammafl_b$ matrices satisfy the well-known Clifford algebra
\begin{equation}
\left\{\gammafl^a,\gammafl^b\right\}=2\,\eta^{ab}\,\Id\:,
\label{eq:Cliffalgflat}
\end{equation}
where $\eta^{ab}$ is the inverse flat Minkowski metric
\begin{equation}
\eta_{ab}=\mathrm{diag}(-1,\,+1,\,+1)\:.
\end{equation}

\par

\paragraph{Curved background.}
In order to take into account the presence of curvature, we consider the customary curved background generalization of \eqref{eq:flatDiracaction} for zero mass fermions \cite{Wald:1984rg,Brill:1966tia,Birrell:1982ix,Cortijo:2006mh,Cortijo:2006xs,deJuan2007charge,
Vozmediano:2010zz,deJuan:2009ldt,deJuan:2012hxm,Amorim:2015bga,Gallerati:2021rtp}: %
\begin{equation}
\mathcal{S}\=i\,\hbar\,\vF \int{d^{3}x\,\sqrt{g}\;\bar{\psi}\,\gamma^\mu\,\DD[\mu]\psi}\;,\qqquad
\label{eq:curvedDiracaction}
\end{equation}
\sloppy
where the curved index $\mu=0,1,2$ refers to the deformed background with metric $\gmat$, while ${\sqrt{g}\equiv\sqrt{-\det(\gmat)}}$ derives from the requirement of a diffeomorphic--covariant form of the new curved spacetime action.%
\footnote{%
This factor, combined with the constant $\vF$, gives rise to sort of space dependent Fermi velocity; the same feature also emerges from a pure tight--binding computation \cite{deJuan2007charge,deJuan:2012hxm,yan2013superlattice,Ghosh:2021yly}
}
\par
The $\gamma_\mu$ matrices for the curved background can be retrieved from the flat--frame counterparts $\gammafl_a$ using the vielbein $\viel$ as:%
\footnote{%
\sloppy
The vielbein defines a local set of tangent frames of the spacetime manifold and is expressed as ${\viel(x)=\frac{\partial y^a (x_0)}{\partial x^\mu}}$, where $y^a(x_0)$ denote a coordinate frame inertial at the space-time point $x_0$.
}
\begingroup%
\begin{equation}
\gamma_\mu\=\viel\,\gammafl_a\:,
\end{equation}
\endgroup
whereas the inverse vielbein $\vielinv$ implements the transformation in the opposite way. The upper indexed gamma matrices are easily obtained through the action of the inverse metric,
\begin{equation}
\gamma^\mu=\gmatinv\,\gamma_\nu\:,
\end{equation}
and satisfy the curved space Clifford algebra:
\begin{equation}
\{\gamma^\mu,\,\gamma^\nu\}\=2\,\gmatinv\,\Id\:.
\label{eq:Cliffalgcurv}
\end{equation}
The diffeomorphic covariant derivative reads
\begingroup%
\begin{equation}
\DD\=\dd+\Omega_\mu\=\dd+\frac{1}{4}\,\spc\,M_{ab}\;,
\label{eq:covder}
\end{equation}
\endgroup
where
\begin{equation}
M_{ab}=\frac12\,[\gammafl_a,\gammafl_b]
\end{equation}
are the Lorentz generators and $\spc$ is the \emph{spin connection}, acting as the gauge field of the local Lorentz group. In a torsionless framework, $\spc$ and $\viel$ are not independent \cite{Eguchi:1980jx,Green:1987sp} and the former can be expressed in terms of the latter as  \cite{Kleinert1989gauge,Katanaev:1992kh}
\begin{equation}
\spc \= \viel[\nu][a]\,\dd\vieluu[\nu][b]+\viel[\nu][a]\,\Conn[\mu][\lambda][\nu]\,\vieluu[\lambda][b]\;,
\end{equation}
where $\Conn$ is the affine connection
\begin{equation}
\Conn= \frac12\,\gmatinv[\sigma][\lambda]
       \left(\dd\gmat[\nu][\sigma]+\dd[\nu]\gmat[\mu][\sigma]-\dd[\sigma]\gmat\right)\;.
\end{equation}
If we consider the above \eqref{eq:covder}, $\Omega_\mu$ plays the role of a gauge field encoding the geometric deformations of the substrate \cite{Semenoff:1984dq,CastroNeto:2009zz}.
The equations of motion for the massless Dirac fermions of action \eqref{eq:curvedDiracaction} are written as%
\footnote{%
from now on we work in (pseudo)natural units: $\hbar=\vF=1$
}
\cite{Birrell:1982ix,Brill:1966tia,Cortijo:2006mh,Cortijo:2006xs,Vozmediano:2010zz,
Gallerati:2018dgm,Gallerati:2021rtp}
\begin{equation}
i\,\gamma^\mu\,\DD\psi\=0\;,
\label{eq:Direqcurvmassless}
\end{equation}
which can be regarded as a generalized form of the flat--space massless Dirac equation. The whole procedure has then led to the substitutions
\begingroup%
\begin{equation}
\begin{split}
\gammafl^a&\:\rightarrow\:\gamma^\mu\;,
\\[2\jot]
\dd[a]&\:\rightarrow\:\DD\:.
\end{split}
\end{equation}
\endgroup
In summary, we have derived a large--wavelength (continuum) formulation for the Dirac modes on the curved graphene--like layer. The curvature effects can be characterized coupling the Dirac spinors to the new $\gmat$ metric, the latter directly affecting the fermion dynamics in the curved background \cite{Cortijo:2006mh,Cortijo:2006xs,Vozmediano:2010zz,deJuan2007charge,Boada:2010sh,
Gallerati:2018dgm,Gallerati:2021rtp}.
In this regime, we can also neglect lattice imperfections if the defects are homogenously distributed and not too much concentrated, to avoid reciprocal interactions.\par
In the following sections we will consider an explicit parametrization of the curved membrane, also providing analytic solutions of the Dirac equation in the related curved background.

\medskip

\subsection{Cylindrical geometry}
Let us consider a two--dimensional surface tightly wrapped around itself. The (cylindrical) symmetry of the space will determine the explicit form of the Dirac equation solutions, while the boundary conditions will lead to a quantization condition on the transverse component of the charge carriers momentum.\par
The parametrization can be written in terms of three--dimensional spacetime coordinates $x^\mu_{{}_{(3)}}$:
\begin{equation}
x^\textsc{a}_{{}_{(4)}}\equiv x^\textsc{a}
=\big(t,\,x,\,y,\,z\big)
\;\quad\longrightarrow\quad\;
x^\mu_{{}_{(3)}}\equiv x^\mu
=\big(t,\,\varphi,\,z\big) \;,
\end{equation}
and, using cylindrical coordinates, one has
\begin{equation}
\begin{cases}
\;x\=r(\varphi)\;\cos(\varphi)\:,
\\[0.75\jot]
\;y\=r(\varphi)\;\sin(\varphi)\:,
\\[0.6\jot]
\;z\=z\;,
\end{cases}
\end{equation}
where we have made explicit the $\varphi$--dependence of the radius. The Jacobian reads
\begin{equation}
J_\mu^{{}^{\,\text{A}}}\=\frac{\partial x^\textsc{a}}{\partial x^\mu}
    \=\left(
        \begin{array}{ccc}
        1  &                      0                                &  0  \\[2\jot]
        0  & -r(\varphi)\,\sin(\varphi)+r'(\varphi)\,\cos(\varphi) &  0  \\[2\jot]
        0  & +r(\varphi)\,\cos(\varphi)+r'(\varphi)\,\sin(\varphi) &  0  \\[2\jot]
        0  &                        0                              &  1  \\[2\jot]
        \end{array}
    \right)  \quad,
\end{equation}
where
$r'(\varphi)\equiv\dfrac{\partial r(\varphi)}{\partial\varphi}$\;.\par\smallskip
The metric $\gmat$ of the curved space has the form:
\begin{equation}
\gmat
    \=\left(J_\mu^{{}^{\,\text{A}}}\right)^{\!\textsc{t}}\eta_\textsc{ab}\;\,J_\nu^{{}^{\,\text{B}}}
    \=
      \left(
        \begin{array}{ccc}
        1   &              0                 &   0    \\[\jot]
        0   & -r(\varphi)^2-{r'(\varphi)}^2  &   0    \\[\jot]
        0   &              0                 &  -1    \\[\jot]
        \end{array}
      \right)  \quad,
\end{equation}
so that the line element reads
\begin{equation}
ds^2\=\gmat\,dx^\mu dx^\nu
    \=dt^2-d\varphi^2\left(r(\varphi)^2+{r'(\varphi)}^2\right)-dz^2\;.
\label{eq:dx2}
\end{equation}
We parameterize the vielbein as
\begin{equation}
\viel \=\left(
        \begin{array}{ccc}
            1  &  0  &  0    \\[\jot]
            0  &  \sqrt{r(\varphi)^2+r'(\varphi)^2}  &  0   \\[\jot]
            0  &  0  &  1    \\[\jot]
        \end{array}
        \right)  \quad,
\label{eq:vielbein}
\end{equation}
and the curved $\gamma_\mu=\viel\,\gammafl_a$ matrices for the cylindrical space then read:
\begin{equation}
\gamma_0=
        \left(\begin{array}{cc}
           1  &  0  \\
           0  & -1  \\
        \end{array}\right)\,,
\qquad
\gamma_1=
        \left(\begin{array}{cc}
           0             &  -i\;f(\varphi)  \\
           -i\;f(\varphi) &       0         \\
        \end{array}\right)\,,
\qquad
\gamma_2=
        \left(\begin{array}{cc}
           0  & -1  \\
           1  &  0  \\
        \end{array}\right)\,,
\label{eq:gammacurv}
\end{equation}
with
\begin{equation}
f(\varphi)\=\sqrt{r(\varphi)^2+{r'(\varphi)}^2}\:.
\label{eq:fphi}
\end{equation}
It is easily demonstrated that the gamma matrices with upper indices ${\gamma^\mu=\gmatinv\,\gamma_\nu}$\, satisfy the Clifford algebra \eqref{eq:Cliffalgcurv}.

\subsubsection{Solution of Dirac equation in cylindrical coordinates}
Having obtained the curved $\gamma$ matrices and having suitably expressed the covariant derivative, we can exploit the Dirac formulation to characterize the electron dynamics in the curved framework.\par
Let us consider the discussed, cylindrically symmetric background. The general curved space Dirac equation reads:%
\footnote{%
Here we consider, for completeness, the generic formulation that takes into account also massive fermions; in the following, we will restrict to the study of Dirac materials featuring massless fermionic charge carriers ($m\to0$ limit).
}
\begin{equation}
(i\,\gamma^\mu\,\mathcal{D}_\mu - m\,\Id)\;\Psi \= 0 \;,
\label{eq:direqcurv}
\end{equation}
where the $\gamma$ matrices are those of eq.\ \eqref{eq:gammacurv}, while the covariant derivative has been defined in \eqref{eq:covder}. The spin connection vanishes as a consequence of the space symmetry, so that the Dirac equation simplifies in:
\begin{equation}
(i\,\gamma^\mu \dd - m\,\Id)\;\Psi \= 0\;.
\label{eq:direqcyl}
\end{equation}
An explicit solution of the above equation reads:
\begin{equation}
\Psi \= e^{-i\,\lambda\,E\,t}\,
        \left(\begin{array}{c}
               \Phi_\textsc{a}(\varphi,z) \\[\jot]
               \Phi_\textsc{b}(\varphi,z) \\[\jot]
        \end{array}\right) \quad,
\label{eq:Psisol}
\end{equation}
where
\begin{equation}
\begin{split}
\Phi_\textsc{a}(\varphi,z)&\=\Cl\;\exp{\!\bigg(i\,k_\varphi\,\int^{\,\varphi}{\!\!f(\varphi')\,d\varphi'}\bigg)}\;\exp{\big(i\,k_z\,z\big)}\;,
\\[1ex] \Phi_\textsc{b}(\varphi,z)&\=\Cl\;\exp{\!\bigg(i\,k_\varphi\,\int^{\,\varphi}{\!\!f(\varphi')\,d\varphi'}\bigg)}\;\exp{\big(i\,k_z\,z\big)}\;\;\frac{-i\,k_\varphi-k_z}{\lambda\,E+m}\;,
\label{eq:PhiABsol}
\end{split}
\end{equation}
with
\begin{equation}
\begin{split}
&\Cl\=\sqrt{\frac{\lambda\,E+m}{2\,E}}\:,
\\[2ex]
&f(\varphi)\=\sqrt{r(\varphi)^2+{r'(\varphi)}^2}\:,\qquad\quad
\\[2ex]
&\lambda\=\pm1\:,
\end{split}
\label{eq:Cnormal}
\end{equation}
and where $\lambda$ labels hamiltonian eigenstates with positive or negative eigenvalues \cite{Gallerati:2018dgm}. The above solution satisfies the Dirac equation \eqref{eq:direqcyl}, once imposed the on-shell condition:
\begin{equation}
\quad
E^2\={\vec{k}}^{\,2}+m^2\:,
\label{eq:onshellcond}
\end{equation}
where $\vec{k}=(k_\varphi,\,k_z)$\,.
\par\medskip

\paragraph{Boundary conditions.}
The periodicity condition on the angular $\varphi$--coordinate of the cylindrical tightly wrapped membrane can be written as
\begin{equation}
\varphi=\varphi+2\pi n\;,\qquad\quad
n\in\mathbb{N}\;.
\end{equation}
Periodic solutions in $\varphi$ requires the exponential term in \eqref{eq:PhiABsol}
\begin{equation}
\exp\bigg(i\,k_\varphi \int^{\,\varphi}{\!\!f(\varphi')\,d\varphi'}\bigg)
    ~\equiv~\exp\big(i\,\vartheta(\varphi)\big)\;,
\end{equation}
to be subject to the boundary condition:
\begin{equation}
\vartheta(2\pi)-\vartheta(0)\=2\pi n\;, \qquad\quad
n\,\in\,\mathds{N}\;.
\end{equation}
This in turn determines the definition of a geometrical parameter $\zeta$,
\begin{equation}
\zeta~\equiv~
\int\limits^{2\pi}_{0}{\!f(\varphi')\,d\varphi'}
\=\frac{2\pi n}{k_\varphi}\;,
\end{equation}
expressing a measure of the cylinder circumference and giving rise to a \emph{quantization condition} for the $k_\varphi$--momentum:
\begin{equation}
k_\varphi \= \frac{2\pi n}{\zeta}\;.
\label{eq:kquantcond}
\end{equation}
\par
In the following sections, we will consider a possible application of the described theoretical framework to Dirac materials, featuring massless charge carriers. In particular, we will apply our geometric tools in the definition of a suitable Kubo formula for curved space configurations.

\bigskip

\section{Optical conductivity}\label{sec:Optical}
Now we consider some detectable properties of a graphene--like curved substrate, whose charge carriers are described by a massless Dirac spectrum ($m=0$). To this end, we will exploit the mathematical tools derived from our analogue--gravity approach. In particular, we will be interested in a suitable definition for the optical conductivity of a graphene--like curved layer.

\subsection{Kubo formula}
Let us start from the Kubo formula for the optical conductivity \cite{kubo1956general,kubo1957statistical,mahan2000many} adapted to a bidimensional background \cite{pereira2011optical,oliva2014anisotropic,chaves2014optical,li2011optical}:
\begin{equation}
\sigma_{\!\mu\nu}\=
i\;\frac{\IeS}{\Omega}\;\sum_{\alpha,\beta}\,\frac{\FFD(E_\alpha-\muchem)-\FFD(E_\beta-\muchem)}{\Omega-\Omega_{\alpha\beta}-i\,\epsilon}\;{v_\mu}^{\!\alpha\beta}\,{v_\nu}^{\!\beta\alpha}\;.
\label{eq:Kubo}
\end{equation}
In the above expression, $\FFD(\Upsilon)$ is the Fermi--Dirac function
\begin{equation}
\FFD(\Upsilon)\=\frac{1}{1+e^{\Upsilon/(k_\textsc{b}T)}}\;,
\end{equation}
\sloppy
$k_\textsc{b}$ is the Boltzmann constant, \,$T$ is the system temperature, \,$\muchem$\, is the chemical potential,\, $\Omega$ is the light energy,\, ${\Omega_{\alpha\beta}=E_\beta-E_\alpha}$ is the transition energy between states $\alpha$ and $\beta$,\, $\IeS=\IeS(e,S)$ is a prefactor depending on the electric charge of the pseudoparticles and layer dimension $S$,\, $\epsilon$ represents an infinitesimal quantity.\par
The ${v_\mu}^{\!\alpha\beta}$ objects express the matrix elements of the velocity operator $\hat{v}_\mu$, defined as:
\begin{equation}
{v_\mu}^{\!\alpha\beta}\=\langle \alpha|\,\hat{v}_\mu\,|\beta\rangle\;.
\label{eq:matrixel}
\end{equation}
The operator $\hat{v}_\mu$ is obtained from the standard quantum mechanical expression
\begin{equation}
\hat{v}_\mu=i\,\left[\hat{H},\,\hat{x}_\mu\right]\;.
\end{equation}
The exact expressions of the operators can be found once given the explicit form of the system Hamiltonian. In Sect.\ \ref{sec:Nanoscrolls} we will explicitly study the case of a cylindrical geometry.
\par\smallskip

\paragraph{Longitudinal conductivity.}
Let us now consider the longitudinal optical conductivity, namely:
\begin{equation}
\sigma_{\!\mu\mu}\=
i\,\frac{\IeS}{\Omega}\:\sum_{\alpha,\beta}\,\frac{\FFD(E_\beta-\muchem)\-\FFD(E_\alpha-\muchem)}{\Omega-\Omega_{\alpha\beta}-i\,\epsilon}\;{\big|{v_\mu}^{\!\alpha\beta}\big|}^2\;.
\label{eq:sigma}
\end{equation}
For infinitesimal values of $\epsilon$, one has
\begin{equation}
\frac{g(x)}{x+i\,\epsilon}\=\frac{g(x)}{x}-i\,\pi\;\delta(x)\:g(x)
\qquad\quad \text{with}\;\;\epsilon\rightarrow0\;,
\end{equation}
and the real part of the optical conductivity reads:
\begin{equation}
\Real[\sigma_{\mu\mu}]\=
\frac{\pi\,\IeS}{\Omega}\,\sum_{\alpha,\beta}\Delta_{\textsc{fd}}\;{\big|{v_\mu}^{\!\alpha\beta}\big|}^2\;\,\delta\!\left(\Omega-\Omega_{\alpha\beta}\right)\,,
\label{eq:Resigma}
\end{equation}
with
\begin{equation}
\Delta_{\textsc{fd}}\:\equiv\:\Delta_{\textsc{fd}}\big(E_\alpha,E_\beta,\muchem\big)
    \=\FFD(E_\beta-\muchem)-\FFD(E_\alpha-\muchem)\;.
\label{eq:Delta}
\end{equation}
The Dirac delta acts as an energy-conservation constraint. The latter, together with the on-shell condition \eqref{eq:onshellcond}, can be used to fix the momentum values. In particular, we find:
\begin{equation}
\delta\big(\Omega-\Omega_{\alpha\beta}(\kdot)\big)~\equiv~
\delta\big(g(\kdot)\big) \=
\sum_{\kdotzero} \frac{\delta\!\left(\kdot-\kdotzero\right)}{\big|g'(\kdotzero)\big|}\;,
\label{eq:DiracDeltaConstr}
\end{equation}
the energy transition $\Omega_{\alpha\beta}$ explicitly depending on the $\kdot$ momentum component. The summation runs over the values $\kdotzero$ that are roots of the Dirac delta argument $g(\kdot)$, that is, the values solving the energy conservation constraint $\Omega=\Omega_{\alpha\beta}$.\par
Putting all together, we obtain an explicit form for the real part of the optical conductivity, reading:
\begin{equation}
\Real[\sigma_{\mu\mu}]\=
\frac{\pi\,\IeS}{\Omega}\:\sum_{\alpha,\beta}\Delta_{\textsc{fd}}\;{\big|{v_\mu}^{\!\alpha\beta}\big|}^2\;\,\WW\;\,\delta\!\left(\kdot-\kdotzero\right)\,, \label{eq:ResigmaD}
\end{equation}
with
\begin{equation}
\WW\=\frac{1}{\left|\dfrac{d\Omega_{\alpha\beta}}{d\kdot}(\kdotzero)\right|}\;.
\label{eq:WOmega}
\end{equation}
%
%
%
In the following section we are going to apply the described formalism to investigate the optical properties of graphene nanoscrolls. This will allow for an easier description of some optical properties of the substrate, taking advantage of the geometrical formulation.

\bigskip

\section{Graphene nanoscrolls}\label{sec:Nanoscrolls}
The richness of optical and electronic properties of graphene--like materials has  attracted a great interest. In particular, high mobility and peculiar optical transitions of the $\pi$ electron bands make this kind of substrate ideal for nanoscale applications (for example, electronic devices, photonic and optoelectronic systems), thanks to its unique optical and electronic properties \cite{stankovich2006graphene,bonaccorso2010graphene,
sun2010graphene,lui2010ultrafast,moradi2020optical}.

\paragraph{Nanoscrolls.}
Graphene nanoscrolls \cite{xie2009controlled,braga2004structure,kim2011multiply} have received great attention due to their unusual properties and potential applications \cite{chen2007structural,mpourmpakis2007carbon,berman2015macroscale,
li2018superior,saini2021low}. They consist in carbon--based structures obtained by rolling a graphene layer into a cylindrical geometry \cite{xie2009controlled}, and exhibit some exciting electronic and mechanical properties due to their distinctively different structure \cite{zhang2011bending,xu2010geometry}, making them conceptually interesting and experimentally relevant.\par
Nanoscrolls formation is dominated by two major energetic contributions, the elastic energy increase caused by bending the graphene sheet (decreasing stability) and the free energy decrease generated by the van der Waals interaction energy
of overlapping regions of the sheet windings (increasing stability). The final structure may have lower energy (higher stability) than the original flat sheet and the scroll formation is a self--sustained curling process, after a critical overlap area is reached \cite{perim2009structure,braga2004structure,saini2021low}.\par
As the bidimensional sheet is rolled up to form the cylindrical structure, its core alters its characteristics, then suggesting a dependency of the nanoscroll properties on its geometrical configuration. Then, nanoscroll electronic features are quite distinct from those of simple flat graphene sheets or carbon nanotubes. In particular, the electronic transport in carbon nanoscrolls is affected by the interlayer $\pi$--$\pi$ windings interactions, also providing structural stability \cite{chen2007structural,pan2005ab}.

\subsection{Cylindrical geometry}
Unusual electronic and optical properties of carbon nanoscrolls are due to their unique topology and distinctive structure \cite{fogler2010effect}.
From a geometrical point of view, nanoscrolls geometry can be parameterized in terms of Archimedean--type spirals in cylindrical coordinates \cite{chen2022formation,trushin2021stability,hassanzadazar2016electrical}.
In order to describe nanoscroll optical properties, we have to elaborate the results of previous Sect.\ \ref{sec:Optical}, taking into account the geometry of the system.

\captionsetup[figure]{skip=5pt,belowskip=20pt}
\begin{figure}[!ht]
\centering
\vspace{0em}
\includegraphics[width=0.56\textwidth,keepaspectratio]{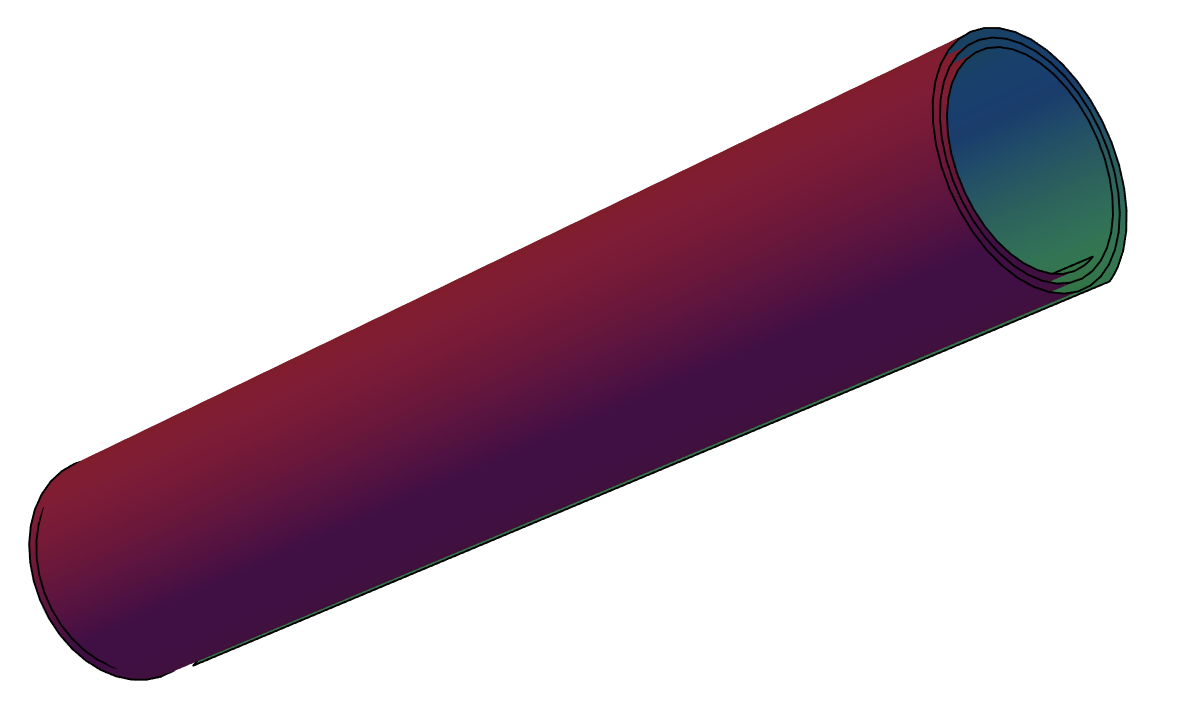}
\caption{\sloppy%
    Nanoscroll topology.}
\label{fig:nanoscroll}
\end{figure}
In particular, in order to find a handy expression for the real part of the optical conductivity, we need the explicit form of the velocity operators matrix element \eqref{eq:matrixel} and of the $\WW$ function of eq.\ \eqref{eq:WOmega}.

\paragraph{Optical conductivity.}
The Hamiltonian of the cylindrical graphene membrane can be obtained from the curved--background Dirac equation. Starting from the time--dependent quantum mechanics equation
\begin{equation}
i\,\partial_t\Psi\=\hat{H}\,\Psi\:.
\end{equation}
Using the explicit expressions of the $\gmatinv$ inverse metric and curved $\gamma_\mu$ matrices, one can isolate the $0$--term of the massless Dirac equation \eqref{eq:direqcyl} obtaining
\begin{equation}
i\,\partial_t\Psi\=i\,\left(\frac{\gamma_0\gamma_1}{f(\varphi)^2}\,\dm[\varphi]+\gamma_0\gamma_2\,\dm[z]\right)\,\Psi\;,
\end{equation}
and the cylindrical--space Hamiltonian operator then reads:
\begin{equation}
\hat{H}_\text{cyl}\=i\,\frac{\gamma_0\gamma_1}{f(\varphi)^2}\,\dm[\varphi]+i\,\gamma_0\gamma_2\,\dm[z]\quad.
\end{equation}
The velocity operator can be the calculated from
\begin{equation}
\hat{v}_\mu\=i\,\left[\hat{H}_\text{cyl}\,,\,\hat{x}_\mu\right]\;,
\end{equation}
giving the explicit expressions
\begin{subeqs} \label{eq:veloper} 
\begin{align}
\hat{v}_\varphi&\=-\frac{\gamma_0\gamma_1}{f(\varphi)^2}\=-\frac{\gammafl_0\gammafl_1}{f(\varphi)}\;,\\[1.5\jot]
\hat{v}_z&\=-\gamma_0\gamma_2\=-\gammafl_0\gammafl_2\;.
\end{align}
\end{subeqs}
%
%
The final expression for the matrix elements
\begin{equation}
{v_\mu}^{\!\alpha\beta}
    \=\langle\,\Psi^{\prime\,\alpha}\,|\,\hat{v}_\mu\,|\,\Psi^\beta\,\rangle
\end{equation}
defines the optical conductivity \eqref{eq:Resigma} and is obtained from the solution $\Psi^\alpha$ of \cref{eq:Psisol,eq:PhiABsol,eq:Cnormal} and velocity operators of \eqref{eq:veloper}.\par
We label the wavefunction solution $\Psi^\alpha$ in terms of the natural number $n$, coming from the quantization condition \eqref{eq:kquantcond}, and Hamiltonian eigenvalue label $\lambda$:
\begin{equation}
\Psi^\alpha~\equiv~\Psi^{(\lambda,n)}\;.
\label{eq:Psilab}
\end{equation}
The matrix element for the velocity operator $v_\varphi$ is given by
\begin{equation}
{v_\varphi}^{\!\alpha\beta}
    \=2\pi\;\Cl\,\Clp\left(\frac{-i\,k_z+k_\varphi}{\lambda\,E}
        +\frac{i\,k_z+k_\varphi'}{\lambda'\,E'}\right)\;\Gphi(n,n')\;\;,
\label{eq:vphiME}
\end{equation}
with
\begin{equation}
\Gphi(n,n')
    \=\int\limits^{2\pi N}_{0} {\!d\varphi\;\;\frac{1}{f(\varphi)}\:\exp\left(i\,\frac{2\pi(n-n')}{\zetaN} \int^{\varphi}\!\!\!d\varphi'f(\varphi')\right)}\quad,
\label{eq:Gphi}
\end{equation}
and
\begin{equation}
\zetaN~\equiv~
\int\limits^{2\pi N}_{0}{\!d\varphi'\:f(\varphi')}\;.
\end{equation}
The number $N$ takes into account the windings of the wrapped graphene layer. From a physical point of view, it acts as a momentum cutoff, correctly restricting the analysis to the long wavelength continuum approximation \cite{Gallerati:2018dgm}. The latter requires that the number $N$ be small when compared with the ratio of the cylinder circumference to the graphene lattice dimension,
\begin{equation}
N\ll \frac{2\pi R}{a}\:.
\end{equation}
To obtain a final expression for the formula \eqref{eq:ResigmaD}, we also have to consider the constraint originating from the presence of the Dirac delta. The solution of the latter constraint, $\Omega=\Omega_{\alpha\beta}(k_z)$, fixes the the $k_z$ component, in the massless formulation, to the value $k_{z_0}$ (see \eqref{eq:DiracDeltaConstr}):
\begingroup
\begin{equation}
k_z \;\;\longrightarrow\;\;\; k_{z_{0}}
    \=\sqrt{\frac{\left(k_\varphi^2-k_\varphi^{\prime\,2}\right)^2
        -2\,\left(k_\varphi^2+k_\varphi^{\prime\,2}\right)\,\Omega^2+\Omega^4}{4\,\Omega^2}}\;\;.
\label{eq:kz0}
\end{equation}
\endgroup
where, for the $k_\varphi$--component, we use the adapted version of \eqref{eq:kquantcond}:
\begin{equation}
k_\varphi\=\frac{2\pi n}{\zetaN}\;,\qquad\;
k_\varphi'\=\frac{2\pi n'}{\zetaN}\;.
\label{eq:kphi}
\end{equation}
The matrix element for the $\hat{v}_z$ operator is obtained similarly as:
\begin{equation}
{v_z}^{\!\alpha\beta}\=
2\pi\;\Cl\,\Clp\left(\frac{i\,k_\varphi+k_z}{\lambda\,E}+\frac{-i\,k_\varphi'+k_z}{\lambda'\,E'}\right)\;
\Gz(n,n')\;\;,
\label{eq:vzME}
\end{equation}
with
\begin{equation}
\Gz(n,n')\=\int\limits^{2\pi N}_{0} {\!d\varphi\;\;\exp\left(i\,\frac{2\pi(n-n')}{\zetaN}
\int^{\varphi}\!\!\!d\varphi'f(\varphi')\right)}\;\;,
\label{eq:Gz}
\end{equation}
where, again, $k_z$ is fixed to the value $k_{z_0}$ and $k_\varphi$ and $k_\varphi'$ comes from eq.\ \eqref{eq:kphi}. The explicit form of the membrane parametrization should be used to determine the functions \,$\Gphi(n,n')$,\, $\Gz(n,n')$\, that modulate the amplitude of the velocity operator matrix elements.\par
Finally, the $\WW$ function of eq.\ \eqref{eq:WOmega}, in the massless case, can be expressed as:
\begin{equation}
\WW\= \left(k_{z_0}\right)^{-1} \left(\frac{1}{\lambda\,E_0}+\frac{1}{\lambda'\,E'_0}\right)^{-1}
\;;\qquad\quad
E_0 ~\equiv~ E\big|_{{}_{k_z = k_{z_0}}}\;\;.\quad
\label{eq:WOm}
\end{equation}
We then find a final, compact expression for the real part of the longitudinal optical conductivity \eqref{eq:Resigma} in cylindrical geometry, reading:
\begin{subeqs} \label{eq:Resigmaphiphizz} 
\begin{align}
\Respp\=
\frac{\pi\,\IeS}{\Omega}\;\sum_{\alpha,\beta}\,\Delta_{\textsc{fd}}\;\WW\;{\big|{v_\varphi}^{\!\alpha\beta}\big|}^2\;,\\[1ex]
\Reszz\=
\frac{\pi\,\IeS}{\Omega}\;\sum_{\alpha,\beta}\,\Delta_{\textsc{fd}}\;\WW\;{\big|{v_z}^{\!\alpha\beta}\big|}^2\;,
\end{align}
\end{subeqs}
\sloppy
where the function $\Delta_{\textsc{fd}}\equiv\Delta_{\textsc{fd}}(E_\alpha,E_\beta,\muchem)$ is defined in eq.\ \eqref{eq:Delta}. As in expression \eqref{eq:Psilab}, the labels \,${\alpha=(\lambda',n')}$\, and \,${\beta=(\lambda,n)}$\, depend on natural numbers \,$n,n'\in [1,N]$\, and Hamiltonian energy eigenvalues ${\lambda,\lambda'=\pm1}$\,.\par\medskip

\subsection{Experimental effects. A worked out model}
Let us now analyse the generic structure of eqs.\ \eqref{eq:Resigmaphiphizz} determining the longitudinal optical conductivity for our wrapped layer in cylindrical geometry.\par\smallskip
The functions $\Gphi$, $\Gz$ are geometrical factors, determined by the manifold topology characterizing the substrate, modulating the amplitude of $\Real[\sigma]$. In the chosen framework, they explicitly depend on the surface parametrization, as well as on the number of windings of the layer.\par
The function $\WW$ of eq.\ \eqref{eq:WOm} features a certain number of poles, coinciding with the energy values $\Omega_0$ for which the $k_{z_0}$ momentum vanishes:
\begingroup
\begin{equation}
k_{z_0}=0  \;\quad\Longrightarrow\quad\;
    \Omega_0=\left|\frac{2\pi(n\pm n')}{\zetaN}\right|\;,\qquad
    n+n'\leq N\;,
\label{eq:poles}
\end{equation}
\endgroup
giving rise to a number of singularities in the optical conductivity, leading to a multiple peak structure. We also note that the singularities arising from the $\WW$ function are partially removed in real samples by the presence of internal disorder, for example imperfections of the substrate and carrier density inhomogeneities; this cuts out the infinities, leaving us with a configuration characterized by multiple but finite peaks.%
\footnote{
Internal disorder of the sample can be simulated, in the calculations, by performing an ensemble averaging over a distribution of the model parameters \cite{hipolito2012enhanced}.
}%
\par\smallskip

\paragraph{Interband and intraband transitions.}
Let us now briefly analyse the expected effects due to different types of state transitions characterizing the optical conductivity .\par
In the case of \emph{intraband transition}, i.e.\ transitions between states with same energy eigenvalues $(\lambda \lambda'=1)$, the factor $\Delta_{\textsc{fd}}$ in general strongly suppresses the value of $\Resintra$, according to Pauli exclusion principle.
In our curved configuration, however, it is possible to have intraband transition between different windings ($n\neq n'$) that is, between different quantum states. In the latter case, $\Resintra$ is still a suppressed quantity, but we can find transition featuring $n\neq n'$ and energy values $\Omega\approx\Omega_0$, then approaching the poles of the $\WW$ function. We then expect a narrow peak structure for $\Resintra$ for small values of $\Omega$, the peaks coinciding with the poles of eq.\ \eqref{eq:poles}.\par
In the case of \emph{interband transition}, that is transition between states with different energy eigenvalues $(\lambda \lambda'=-1)$, the matrix element ${\big|v_\varphi^{\text{inter}}\big|}$ vanishes for $\Omega=\Omega_0$. In particular, we also have:
\begin{equation}
\lim_{\Omega \to \Omega_0} \,{\left|v_\varphi^{\text{inter}}(\Omega)\right|}^2\;\WW\=0\;,
\label{eq:interbandzeros}
\end{equation}
so that the poles of the $\WW$ function are cancelled by the zeros of the matrix elements, giving rise to a minimum of \,$\Resinterpp$\, for \,$\Omega\to\Omega_0$\,. The optical conductivity is then expected to reflect the quantization condition \eqref{eq:kphi} of the $k_\varphi$--momentum, featuring a certain number of oscillation related to the number $N$ (i.e.\ to the curved sample geometry).
On the other hand, the matrix element ${\big|v_z^{\text{inter}}\big|}$ is different from zero for $\Omega=\Omega_0$, and there is no cancellation of the poles originating from $\WW$: this suggests again the presence of a strongly peaked behaviour for $\Resinterzz$.
\par\smallskip
Clearly, in both interband and intraband case, the conductivity is suppressed for very large values of $\Omega$. We will directly verify the above considerations in the next section, where we will provide an explicit parametrization of the surface and consider suitable values of the other involved physical parameters.

\subsubsection{Explicit parametrization and simulations}
Let us write an explicit parametrization for the cylindrical geometry of our layer surface as:%
\footnote{%
Here we want to exploit our formalism in a non-trivial Archimedean-like parametrization, where also the derivative $r'(\varphi)$ is non-constant, see also \cref{eq:vielbein,eq:gammacurv,eq:fphi}.
}
\begin{equation}
r(\varphi)\=R\,{\left(1-\frac{1}{2\pi}\,\frac{\varphi}{D N}\right)}^{\!2}\;,
\end{equation}
where the coefficient $D$ controls the distance between the layers and $R$ is a typical dimension for the average radius of the cylinder.\par
\begin{figure}[!ht]
\centering
\includegraphics[width=0.75\textwidth]{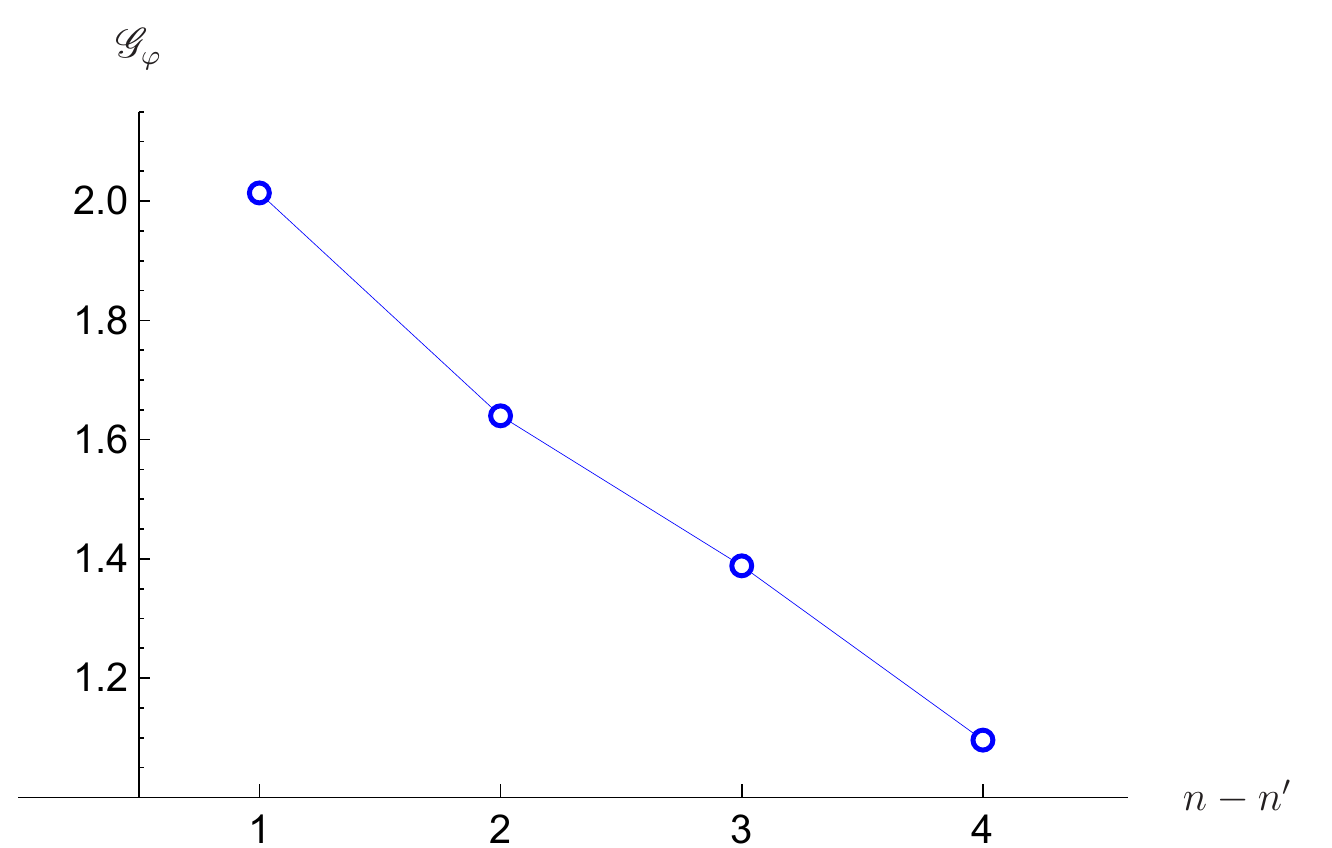}
\caption{Example of the \,$\Gphi(n,n')$\, function for $N=5,\,D=120$\,.}
\label{fig:GGphi}
\end{figure}
\begin{figure}[!ht]
\centering
\includegraphics[width=0.75\textwidth]{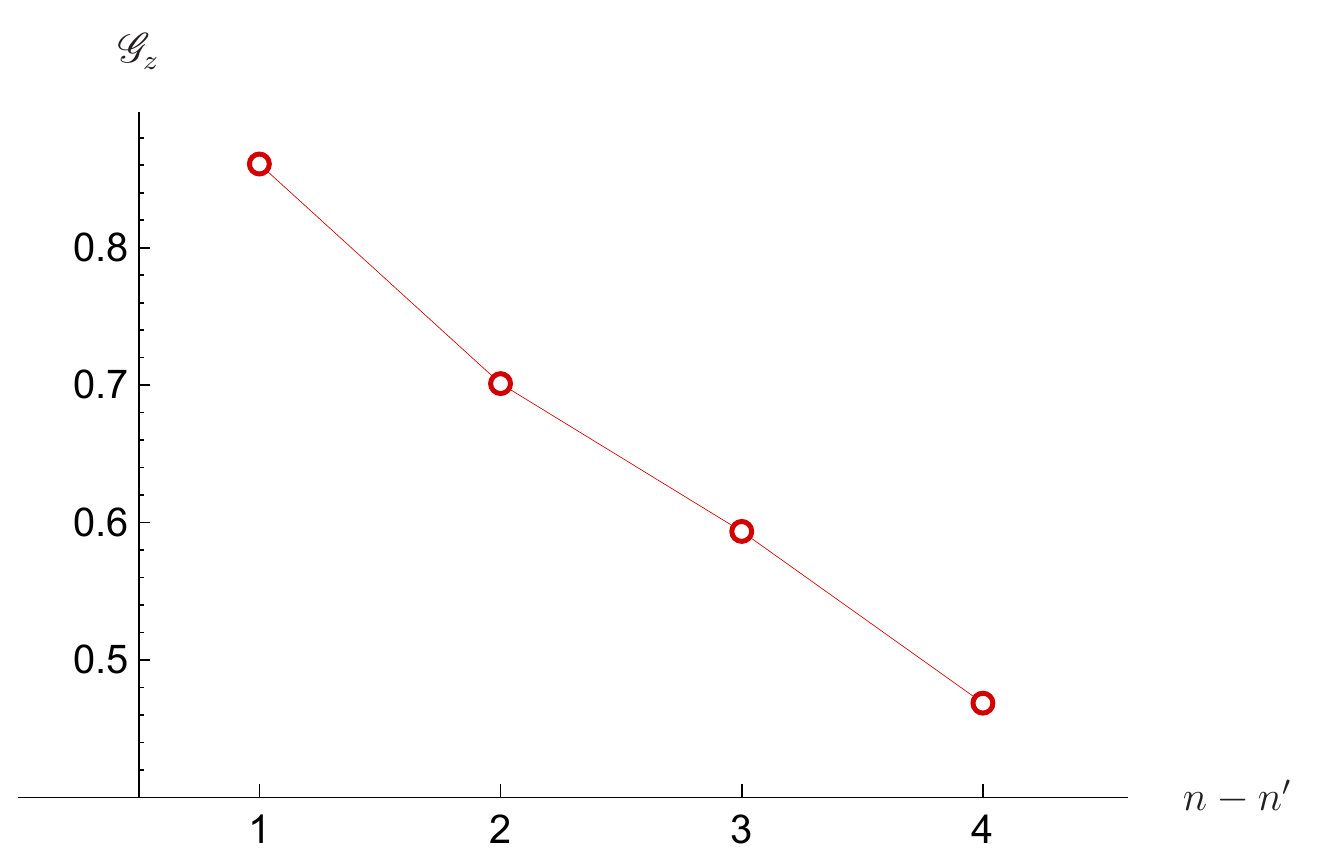}
\caption{Example of the \,$\Gz(n,n')$\, function for $N=5,\,D=120$\,.}
\label{fig:GGz}
\end{figure}

\captionsetup[figure]{skip=5pt,belowskip=20pt}
\begin{figure}[!ht]
\centering
\vspace{0em}
\includegraphics[width=0.75\textwidth,keepaspectratio]{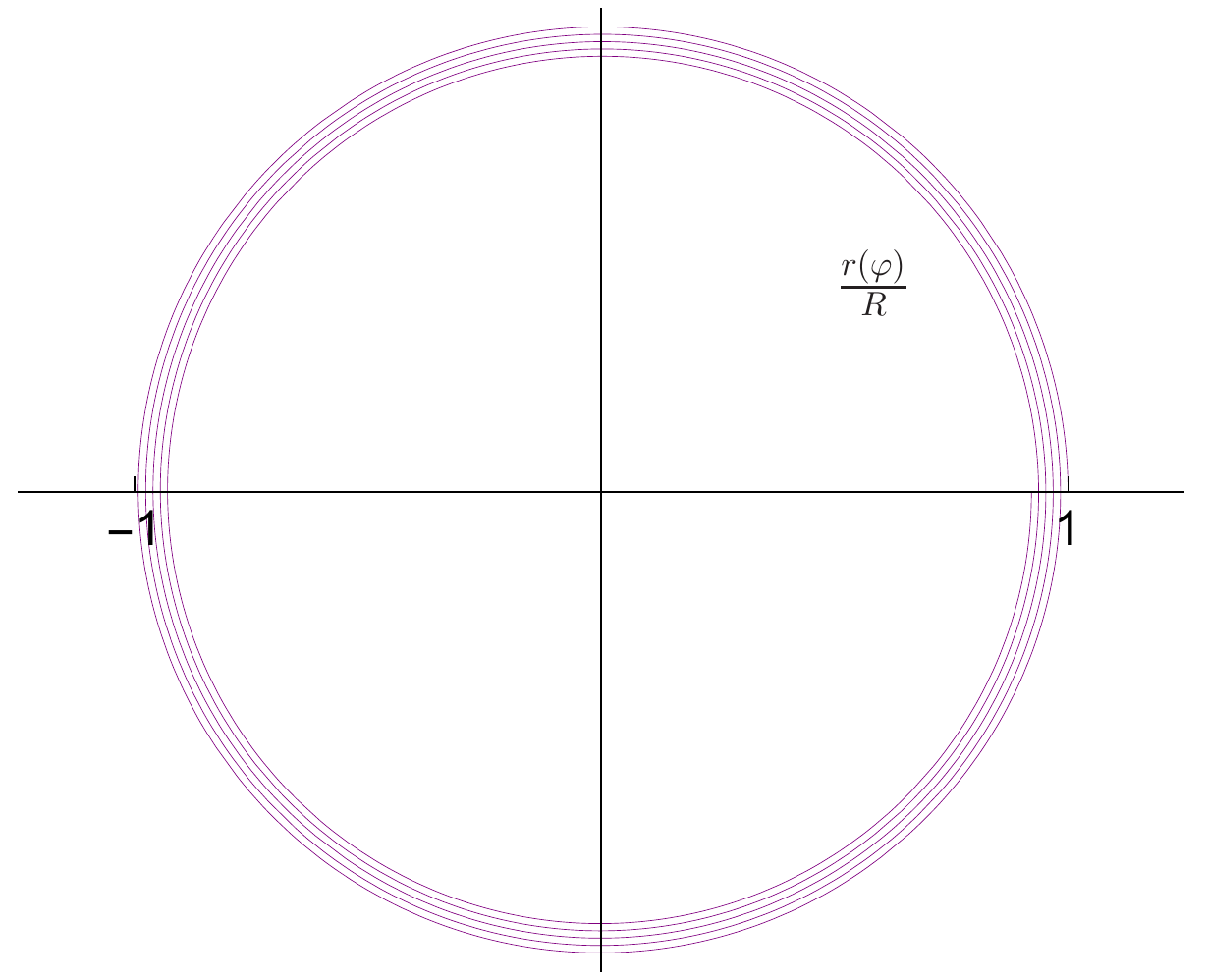}
\caption{Parametric plot for \,${r(\varphi)/R}$,\, with \,$D=25$\, and $N=5$\,.}
\label{fig:spiral}
\end{figure}
The functions $\Gphi$ and $\Gz$ can be obtained using \cref{eq:Gphi,eq:Gz}, depending on the number of windings $N$ and parameter $D$. In \cref{fig:GGphi,fig:GGz}, we plot them for $N=5$ and $D=120$.\par
The poles of the $\WW$ function come from \eqref{eq:poles}. In particular, if we choose $R=90\,\mathrm{nm}$, $N=5$ and $D=120$, we find:
\begin{equation}
\Omega_0\=q\,\times\,0.442\,\mathrm{eV}\;, \;\qquad q=1,2,3,\dots,10\,.
\label{eq:polesN}
\end{equation}
We expect unconventional optical responses from the nanoscroll sample in a range of energy $[0.4\div5]$ eV. Remarkably, the latter energy interval includes the visible light energy range.\par
\begin{figure}[!h]
\vspace{0em}
\centering
\includegraphics[width=0.75\textwidth]{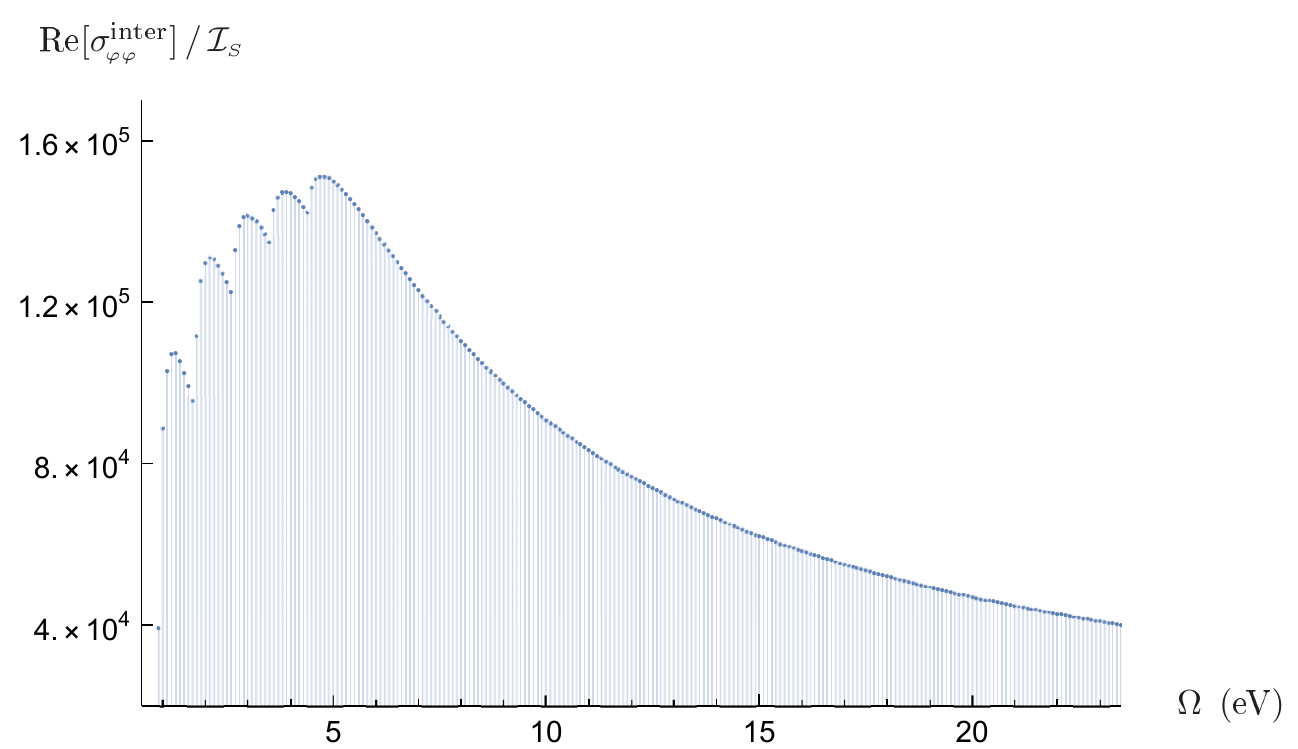}
\caption{Real part of the interband longitudinal optical conductivity $\Resinterpp$ for a nanoscroll geometry with parameters $N=5$,\, $D=120$,\, $R=90\,\nm$,\, $T=300\,\Kelv$,\, $\muchem=0.2\,\eV$.}
\label{fig:ResinterppN5}
\end{figure}

\begin{figure}[!h]
\centering
\includegraphics[width=0.75\textwidth]{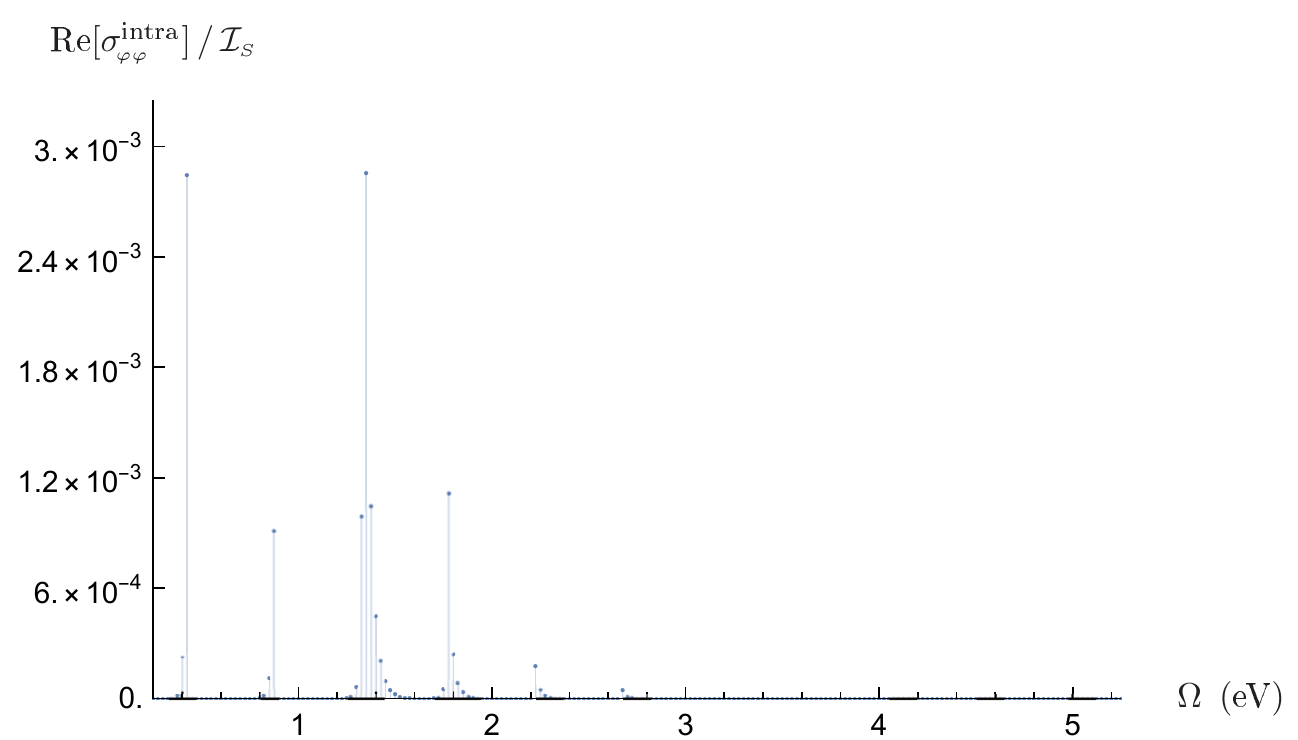}
\caption{Real part of the intraband longitudinal optical conductivity $\Resintrapp$ for a nanoscroll geometry with parameters $N=5$,\, $D=120$,\, $R=90\,\nm$,\, $T=300\,\Kelv$,\, $\muchem=0.2\,\eV$.}
\label{fig:ResintrappN5}
\end{figure}

\begin{figure}[!h]
\vspace{0em}
\centering
\includegraphics[width=0.75\textwidth]{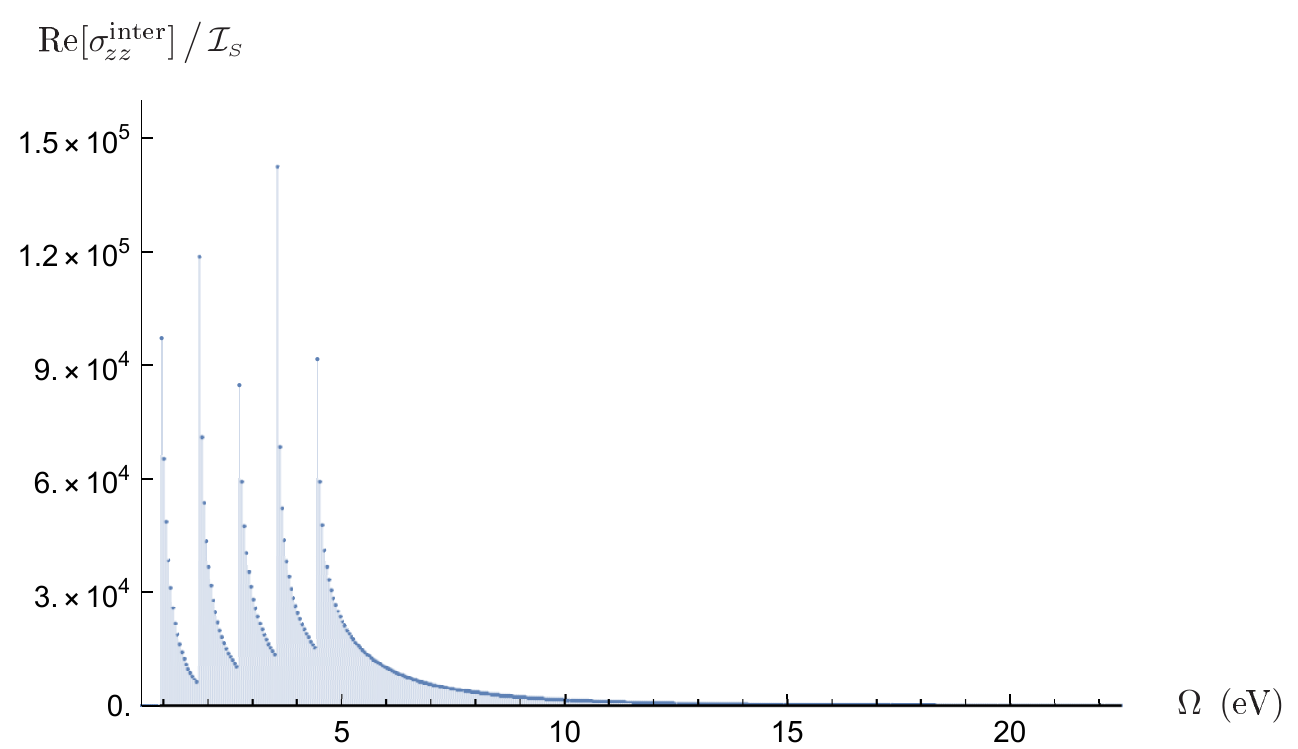}
\caption{Real part of the interband longitudinal optical conductivity $\Resinterzz$ for a nanoscroll geometry with parameters $N=5$,\, $D=120$,\, $R=90\,\nm$,\, $T=300\,\Kelv$,\, $\muchem=0.2\,\eV$.}
\label{fig:ResinterzzN5}
\end{figure}

\begin{figure}[!h]
\vspace{0em}
\centering
\includegraphics[width=0.75\textwidth]{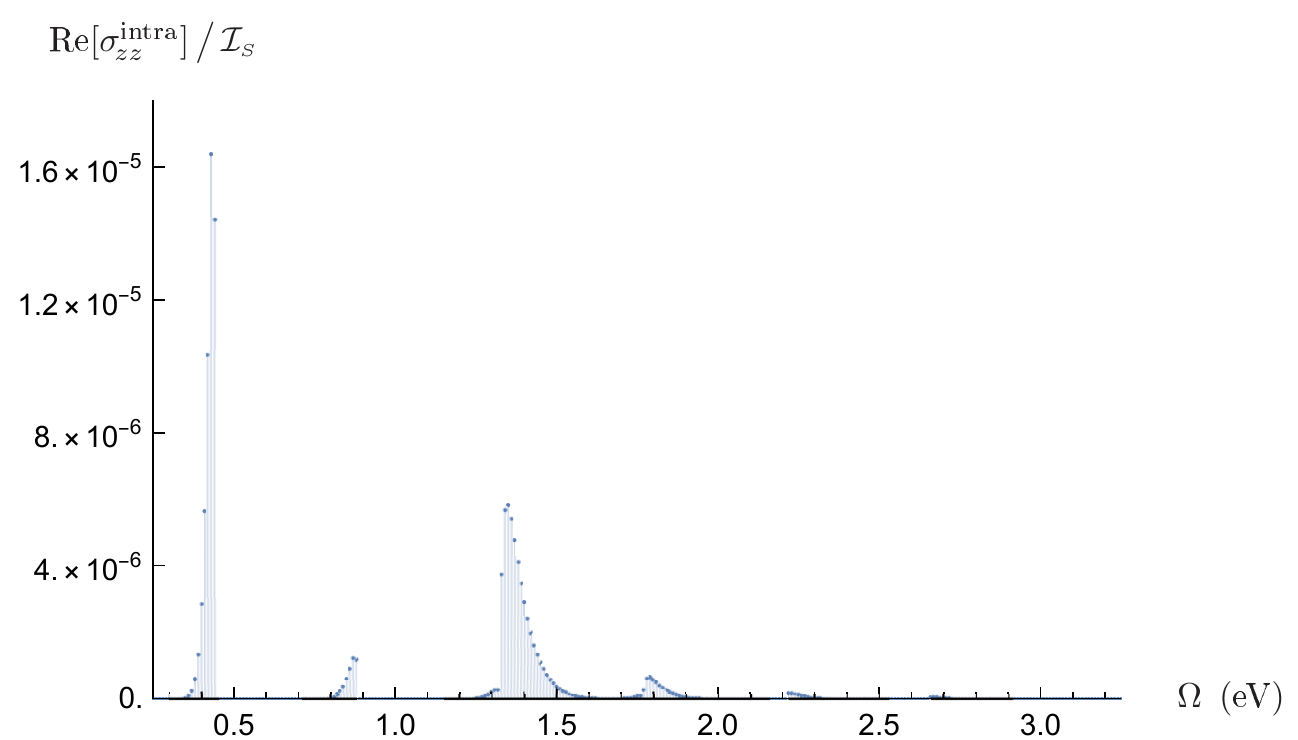}
\caption{Real part of the intraband longitudinal optical conductivity $\Resintrazz$ for a nanoscroll geometry with parameters $N=5$,\, $D=120$,\, $R=90\,\nm$,\, $T=300\,\Kelv$,\, $\muchem=0.2\,\eV$.}
\label{fig:ResintrazzN5}
\end{figure}

\begin{figure}[!h]
\vspace{0em}
\centering
\includegraphics[width=0.75\textwidth]{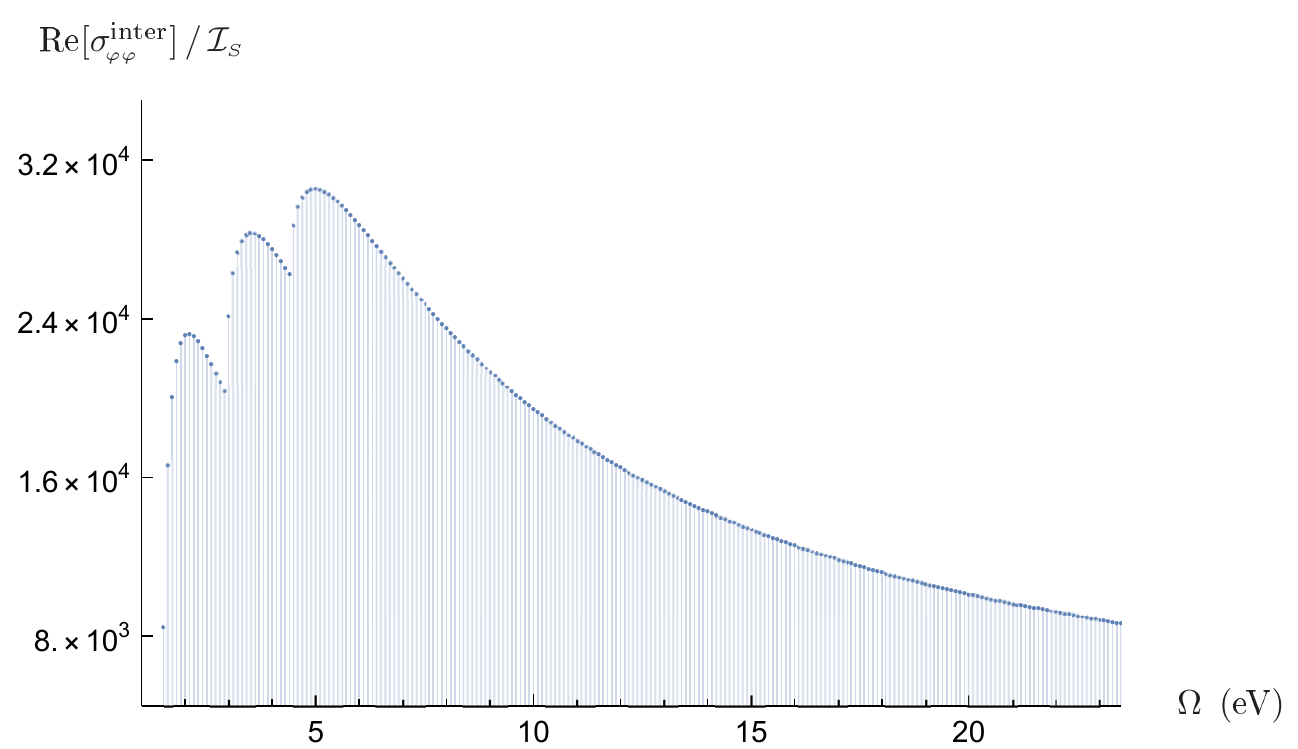}
\caption{Real part of the interband longitudinal optical conductivity $\Resinterpp$ for a nanoscroll geometry with parameters $N=3$,\, $D=120$,\, $R=90\,\nm$,\, $T=300\,\Kelv$,\, $\muchem=0.2\,\eV$.}
\label{fig:ResinterppN3}
\end{figure}

\begin{figure}[!h]
\vspace{0em}
\centering
\includegraphics[width=0.75\textwidth]{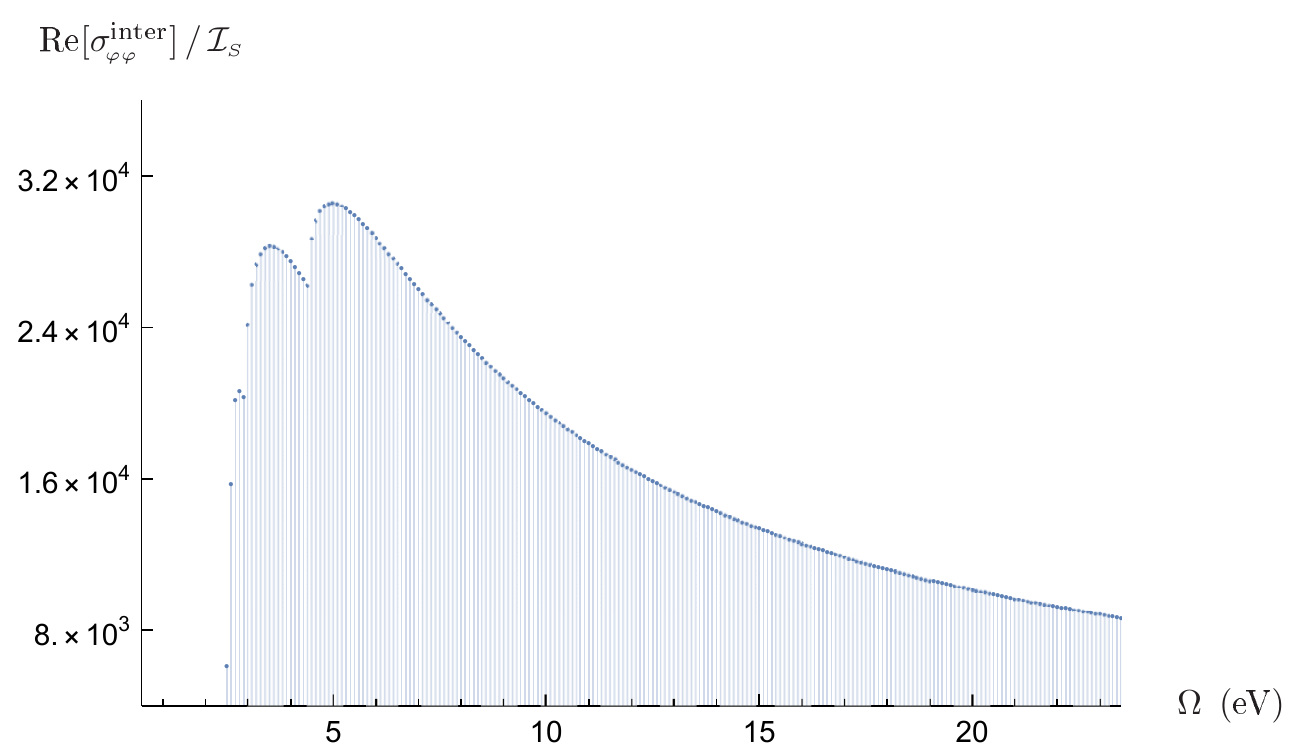}
\caption{Real part of the interband longitudinal optical conductivity $\Resinterpp$ for a nanoscroll geometry with parameters $N=3$,\, $D=120$,\, $R=90\,\nm$,\, $T=300\,\Kelv$,\, $\muchem=1.2\,\eV$.}
\label{fig:ResinterppN3mu2}
\end{figure}

\smallskip
In \cref{fig:ResinterppN5,fig:ResintrappN5,fig:ResinterzzN5,fig:ResintrazzN5} we plot the real part of the longitudinal optical conductivity $\Real[\sigma_{\mu\mu}]$ as a function of the energy $\Omega$, for interband and intraband transitions of a nanoscroll geometry with number of windings $N=5$. As expected from our analysis, the intraband transitions are in general suppressed, except for narrow peaks around the singular points \eqref{eq:polesN}. The interband transitions, on the contrary, show optical responses for the expected energy ranges. The zeros of the interband transitions exactly occur for the points identified by \eqref{eq:interbandzeros}. In fig.\ \ref{fig:ResinterppN3} we plot $\Resinterpp$ for interband transitions with $N=3$. For the latter number of windings, the intraband transition features strongly suppressed values of the optical conductivity ($\Resinterpp/\IeS$ has isolated peaks with maximum values of the order of $10^{-10}$ or less). Finally, in fig.\ \ref{fig:ResinterppN3mu2} we plot $\Resinterpp$ for the same parameters but a different chemical potential.%
\footnote{%
The chemical potential of doped graphene can be controlled by chemical doping or external gate voltage \cite{bao2012graphene}.
}
We can see that this requires larger minimum energy to observe optical responses: in particular, the critical incident light energy triggering the transition increases as the chemical potential rises \cite{zhao2020tunable}.

The massless formulation for charge carriers in graphene nanoscrolls has been recently studied in \cite{chen2022formation}, the gapless nature of the material being preserved especially for high overlap ratio (as it happens for our multiple wrapped surfaces).

\section{Concluding remarks}
A deeper intertwining of different scientific areas has always proved to be a powerful tool for improving our understanding of many fascinating physical aspects of our world.
In particular, by intersecting outcomes from condensed matter and high energy physics, many developments can be found in a multidisciplinary environment, see e.g.\ \cite{Zurek:1996sj,volovik1990superfluid,Volovik:2000ua,
Baeuerle:1996zz,Ruutu:1995qz,Garay:1999sk,
Carusotto:2008ep,Mannarelli:2008jq,Iorio:2010pv,Stegmann:2015mjp,Ummarino:2017bvz,
Capozziello:2018mqy,zaanen2015holographic,Sepehri:2016nuv,Franz:2018cqi,
Ummarino:2019cvw,morresi2020exploring,Capozziello:2020ncr,Gallerati:2020tyq,
Bagchi:2021avw,silva2021intrinsic,Ummarino:2020loo,Gallerati:2021ops,aghaei2021Dirac,
fazlali2020nonlinear,Ummarino:2021vwc,Kolobov:2021ynv,Ummarino:2021tpz,Gallerati:2022pgh}.
Inspired by these approaches, we combined techniques from general relativity, differential geometry and solid state physics to analyse some interesting properties of quantum modes in a concrete, two--dimensional curved background.
\par
The study of graphene--like materials naturally involves a relativistic quantum field formulation, these special materials realizing the physics of Dirac fermions in a real laboratory system. As we have seen, the presence of curvature may give rise to peculiar physical consequences, that can be interpreted as the analogues of (lower--dimensional) gravitational effects, the deformed layer acting as a curved spacetime. In this paper, in particular, we constructed analytical solution to the Dirac equation in a deformed framework representing a cylindrical nanoscroll topology. We then introduced a suitable Kubo formula for a bidimensional background, deriving a compact form for the real part of the optical conductivity. The latter was exploited to obtain some remarkable experimental predictions about the optical response of the nanoscroll in the visible--light energy range at room temperature $T=300\,\mathrm{K}$. Clearly, the same approach finds application also in different spacetime geometries, like for instance the BTZ geometry \cite{Banados:1992wn} that, under certain conditions, may be reproduced in suitable graphene configurations with constant negative curvature \cite{Cvetic:2012vg,Kandemir:2019zyt}.\par
The results have been obtained exploiting a \emph{bottom--up} formulation, the geometry of the graphene--like substrate affecting the propagation of the quantum fields in the material. The dynamics of the charge carriers therefore depends on an effective metric, so that it is possible to exploit some mathematical tools from General Relativity and quantum field theory in curved spacetime.
On the other hand, some studies theorize that graphene--like membranes could realize alternative (unconventional) forms of supersymmetry \cite{Alvarez:2011gd,Alvarez:2013tga,Alvarez:2020qmy,Alvarez:2021qbu,Alvarez:2021zhh}, the latter becoming in turn a new tool to inspect the properties of quantum fields in Dirac materials. This kind of construction exploits an holographic \emph{top--down} approach, where the substrate description originates from an higher--dimensional geometrical formulation of a (super)gravity theory \cite{Andrianopoli:2019sip,Gallerati:2021htm}. A more detailed comparison between these two kinds of approach will be pursued in a future work.

\bigskip

\section*{\normalsize Acknowledgments}
\vspace{-2.5pt}
I would like to thank Profs.\ M.\ Trigiante, G.\ A.\ Ummarino, F.\ Laviano and A.\ Carbone
that supported these studies with their funds.\par\bigskip
%


\newpage

\appendix
\addtocontents{toc}{\protect\setcounter{tocdepth}{1}}
\addtocontents{toc}{\protect\addvspace{3.5pt}}%
\numberwithin{equation}{section}%
\numberwithin{figure}{section}%

\section{Conventions}\label{app:Conventions}
\paragraph{Dirac equation.}
The Dirac equation in flat Minkowski spacetime expresses a relativistic field equation, whose squared wavefunction modulus could be consistently interpreted as a probability density. The latter requirements give rise to a first-order equation in time--derivative, while relativistic invariance fixes the equation to be first order in space-derivatives too. The final explicit form must also fulfill the requests of Lorentz covariance and satisfy the Klein--Gordon equation. The resulting equation reads
\begin{equation}
(i\,\gammafl^a \dd[a] - m\,\Id)\;\psi(x)\=0\;,
\label{eq:Diraceqflatapp}
\end{equation}
together with the condition
\begin{equation}
\left\{\gammafl^a,\,\gammafl^b\right\}\=2\,\eta^{ab}\,\Id\:,
\label{eq:Cliffalgflatapp}
\end{equation}
that is usually referred to as Clifford algebra, $\eta^{ab}$ being the inverse of the Minkowski flat metric $\eta_{ab}\,$.\par
For the sake of notational simplicity, in eq.\ \eqref{eq:Diraceqflatapp} we have omitted the spinorial indices of $\psi\equiv\psi^\beta$\, and \,$\gammafl^a={(\gammafl^a)^\alpha}_\beta$\,. In particular, spinorial quantities behave as scalars under general space-time coordinate transformations, while they transform in a spinor representation $\mathscr{R}$ under the local Lorentz group:
\begingroup
\begin{equation}
\psi^{\prime\,\alpha}(x) \=
{\mathscr{R}^{\,\alpha}{}_{\beta}}\big[\Lambda(x)\big]\;\psi^\beta(x)\:.
\end{equation}
\endgroup
Using the explicit form of the Lorentz generators to construct the Pauli-Lubanski operator, it can be easily shown that the particle has spin $s=\tfrac12$\,.

\subsection{Curved spaces} \label{subapp:curvspace}
In order to conveniently describe curved spaces (i.e., general relativity scenarios) in the presence of spinorial fields, we introduce suitable tools able to describe transformation rules generalized to curved backgrounds.

\paragraph{Vielbein formalism.}
Let us consider a set of coordinates locally inertial, so that one can apply the usual Lorentz spinor behaviour, and let us imagine to find a way to translate back to the original coordinate frame. More precisely, let $y^a(x_0)$ denote a coordinate frame that is inertial at the space-time point $x_0$: we shall call these the ``Lorentz'' coordinates. Then,
\begin{equation}
\viel(x) \= \frac{\partial y^a (x_0)}{\partial x^\mu}
\end{equation}
gives the so-called \emph{vielbein}. It defines a local set of tangent frames of the spacetime manifold and, under general coordinate transformations, it transforms covariantly as
\begin{equation}
{e'_\mu}^{\!a}(x') \=
\frac{\partial x^\nu}{\partial x'^\mu}\:\,\viel[\nu][a](x)\;,
\end{equation}
while a Lorentz transformation leads to
\begin{equation}
{e'_\mu}^a (x) \= {\Lambda^{{\!}^\ml{a}}}_b\:\,\viel[\mu][b](x)\;.
\end{equation}
The space-time metric, in particular, is expressed as
\begin{equation}
g_{\mu\nu}(x) \= \viel[\mu][a](x)\;\viel[\nu][b](x)\;\eta_{ab}\;,
\end{equation}
in terms of the Minkowski flat metric $\eta_{ab}$\,.
The original constant $\gammafl_a$ matrices of the inertial frame can be converted into the new $\gamma_\mu$ matrices of the curved background by the action of the vielbein:
\begin{equation}
\gamma_\mu(x)\=\viel(x)\:\gammafl_a\;,
\end{equation}
while the inverse vielbein $\vielinv$ performs the transformation in the other direction. The vielbein thus takes Lorentz (flat) latin indices to coordinate basis (curved) greek indices. The gamma matrices with upper indices
\begin{equation}
\gamma^\mu\=\gmatinv\,\gamma_\nu\;,
\end{equation}
satisfy the relation:
\begin{equation}
\{\gamma^\mu,\,\gamma^\nu\}\=2\,g^{\mu\nu}\,\Id\;,
\end{equation}
that holds in curved backgrounds and is the equivalent form of the previous, flat Clifford algebra \eqref{eq:Cliffalgflatapp}.

\paragraph{Covariant derivative, spin connection.}
The choice of the locally inertial frame $y^a$ is defined up to Lorentz transformations determined by the Lorentz generators $M_{ab}$\,. In order to couple fields, we define the covariant derivatives:
\begin{equation}
\mathcal{D}_\mu \= \dd + \frac14\,\spc\,M_{ab}\;,
\end{equation}
where
\begin{equation}
M_{ab} \= \frac12\,\left[\gammafl_a,\gammafl_b\right]
\end{equation}
are the Lorentz generators. The $\spc$ object defines the \emph{spin connection}, that can be seen as the gauge field of the local Lorentz group, the corresponding field strength given by the Riemann curvature tensor. It is determined through the vielbein postulate (tetrad covariantly constant):
\begin{equation}
\mathcal{D}_\mu \viel[\nu] - \Conn\,\viel[\lambda] \= 0 \;.
\end{equation}
The latter is expressed in terms of the affine connection $\Conn$
\begin{equation}
\Conn \= \frac12\,\gmatinv[\sigma][\lambda]
       \left(\dd\gmat[\nu][\sigma]+\dd[\nu]\gmat[\mu][\sigma]-\dd[\sigma]\gmat\right)\;.
\end{equation}
while the explicit expression for the spin connection reads
\begin{equation}
\spc \= \viel[\nu][a]\,\dd\vieluu[\nu][b]+\viel[\nu][a]\,\Conn[\mu][\lambda][\nu]\,\vieluu[\lambda][b]\;.
\end{equation}
Finally, the Dirac equation in curved spacetime can be written as:
\begin{equation}
\left(i\,\gamma^\mu\,\DD - m\,\Id\right)\;\psi \= 0 \;.
\end{equation}


\newpage

\hypersetup{linkcolor=blue}
\phantomsection 
\addtocontents{toc}{\protect\addvspace{4.5pt}}
\addcontentsline{toc}{section}{References} 
\bibliographystyle{mybibstyle}
\bibliography{bibliografia} 

\end{document}

%% file: Structure.tex
\usepackage[utf8]{inputenc} 
\usepackage{lipsum} 

\usepackage[top=3.25cm, bottom=3cm, left=3.5cm, right=3.5cm,
           heightrounded,marginparwidth=1.5cm, marginparsep=1cm]{geometry} 

\usepackage{changepage} 

\usepackage[shortlabels]{enumitem} 

\usepackage[square,numbers,merge,comma,sort&compress]{natbib} 
\makeatletter
\def\NAT@spacechar{\,}  
\makeatother

\usepackage{amsmath,amssymb,amsfonts,amsthm,amsbsy}
\usepackage{mathtools} 

\usepackage{fnpct}
\setfnpct{dont-mess-around} 
\usepackage{slashed,cancel}
\usepackage{old-arrows} 
\usepackage{comment}   
\usepackage{relsize}   
\usepackage{setspace}  
\usepackage{moresize}  
\usepackage{epsfig}
\usepackage{latexsym}
\usepackage{mathrsfs,calligra,aurical,bold-extra} 
\usepackage{calc}
\usepackage{float}
\usepackage{appendix}
\usepackage{bigints}
\usepackage{xargs}
\usepackage{extarrows}
\usepackage{empheq}
\usepackage{soul} 

\usepackage[svgnames]{xcolor}  
\definecolor{Green}{rgb}{0.05, 0.45, 0.25}
\xdefinecolor{dogwoodrose}{rgb}{0.8, 0.1, 0.55}
\xdefinecolor{RRed}{rgb}{0.7, 0.1, 0.525}

\usepackage{tikz}
\usetikzlibrary{scopes,decorations.pathmorphing,patterns,calc,arrows,
                shapes.geometric,shapes.arrows,decorations.markings,plotmarks}
\usepackage{pgfplots}

\usepackage[blocks]{authblk}  

\usepackage[fulladjust]{marginnote}%

\usepackage{graphicx} 
\graphicspath{{Figures/}} 

\usepackage[labelsep=colon]{caption}  
\usepackage[labelsep=colon,aboveskip=10pt,belowskip=10pt]{subcaption}
\DeclareCaptionSubType * [roman]{table}
\usepackage{array,multirow,makecell,booktabs}  
\newcolumntype{M}[2]{>{\centering\arraybackslash$}#1{#2\linewidth}<{$}}
\newcolumntype{T}[2]{>{\centering\arraybackslash}#1{#2\linewidth}<{}}
\newcolumntype{R}[1]{>{\raggedleft\arraybackslash}m{#1\linewidth}<{}}
\newcolumntype{L}[1]{>{\raggedright\arraybackslash}m{#1\linewidth}<{}}

\makeatletter
\renewcommand\mcell@classz{\@classx
   \@tempcnta \count@
   \prepnext@tok
   \@addtopreamble{
      \ifcase\@chnum
         \hfil
         \mcell@agape{\d@llarbegin\insert@column\d@llarend}\hfil \or
         \hskip1sp
         \mcell@agape{\d@llarbegin\insert@column\d@llarend}\hfil \or
         \hfil\hskip1sp
         \mcell@agape{\d@llarbegin \insert@column\d@llarend}\or
         \mcell@agape{$\vcenter
         \@startpbox{\@nextchar}\insert@column\@endpbox$}\or
         \mcell@agape{\vtop
         \@startpbox{\@nextchar}\insert@column\@endpbox}\or
         \mcell@agape{\vbox
         \@startpbox{\@nextchar}\insert@column\@endpbox}%
      \fi
      \global\let\mcell@left\relax\global\let\mcell@right\relax
    }\prepnext@tok}
\makeatother

\usepackage{titlesec}

\titleformat{\section}{\normalfont\fontsize{12.5}{12}\bfseries}{\thesection}{0.5em}{}
\titleformat{\subsection}{\normalfont\fontsize{10.5}{10}\bfseries}{\thesubsection}{0.5em}{}
\titleformat{\subsubsection}{\normalfont\normalsize\bfseries}{\thesubsubsection}{0.5em}{}

\titlespacing*{\section}{0pt}%
                {4ex plus 1ex minus .5ex}{1.75ex plus .25ex minus .25ex}
\titlespacing*{\subsection}{0pt}%
                {3.5ex plus 1ex minus .5ex}{1.25ex plus .2ex minus .2ex}
\titlespacing*{\subsubsection}{0pt}%
                {2.5ex plus 0.75ex minus .2ex}{0.75ex plus .15ex minus .15ex}
\titlespacing*{\paragraph}{0pt}%
                {1.85ex plus 0.5ex minus .15ex}{1em}

\usepackage{titletoc}

\addtocontents{toc}{\addvspace{-0.75em}}  

\titlecontents{section}
  [1.25em] {\addvspace{0.7em plus 0pt}\small}
  {\thecontentslabel\hspace{0.75em}}{}
  {\hspace{0.5em}\titlerule*[0.5em]{.}\contentspage}
  [\addvspace{0.0em plus 0pt}]

\titlecontents{subsection}
  [2.75em] {\addvspace{0.075em plus0pt}\fns}
  {\thecontentslabel\hspace{0.75em}}{\thecontentslabel\hspace{0.75em}}
  {\hspace{0.5em}\titlerule*[0.5em]{.}\small\contentspage}
  [\addvspace{0.075em plus 0pt}]

\usepackage{environ}
\makeatletter
\NewEnviron{subalign}[1]{
\begin{subequations}\label{#1}
\begin{align} \BODY \end{align}
\end{subequations}      }
\makeatother
\makeatletter
\newenvironment{subeqs}%
{\begingroup%
\setlength{\abovedisplayskip}{10pt plus 4pt minus 9pt}%
\setlength{\abovedisplayshortskip}{0pt plus 2pt minus 2pt}%
\setlength{\belowdisplayskip}{12pt plus 3pt minus 9pt}%
\setlength{\belowdisplayshortskip}{7pt plus 3pt minus 4pt}%
\begin{subequations}%
}%
{\end{subequations}\ignorespacesafterend%
\endgroup}%
\makeatother

\makeatletter
{\begingroup%
\setlength{\abovedisplayskip}{{#1}pt plus 3pt minus 3pt}%
\setlength{\abovedisplayshortskip}{{#2}pt plus 3pt}%
\setlength{\belowdisplayskip}{{#3}pt plus 3pt minus 3pt}%
\setlength{\belowdisplayshortskip}{{#4}pt plus 3pt minus 3pt}%
\begin{subequations}%
}%
{\end{subequations}\ignorespacesafterend%
\endgroup}%
\makeatother

\makeatletter
{\begingroup%
\setlength{\abovedisplayskip}{{#1}pt plus 3pt minus 3pt}%
\setlength{\abovedisplayshortskip}{{#2}pt plus 3pt}%
\setlength{\belowdisplayskip}{{#3}pt plus 3pt minus 3pt}%
\setlength{\belowdisplayshortskip}{{#4}pt plus 3pt minus 3pt}%
\begin{equation}%
}%
{\end{equation}\ignorespacesafterend%
\endgroup}%
\makeatother

\makeatletter
{%
\begin{equation}%
\begin{split}%
}%
{\end{split}%
\end{equation}\ignorespacesafterend%
}%
\makeatother

\usepackage{bm}  
\usepackage{dsfont}  

\DeclareMathAlphabet{\mathpzc}{OT1}{pzc}{m}{it}
\DeclareMathAlphabet{\mathcal}{OMS}{cmsy}{m}{n}
\DeclareSymbolFontAlphabet{\Scr}{rsfs}
\DeclareMathAlphabet{\mathbold}{U}{BOONDOX-ds}{m}{n}
\SetMathAlphabet{\mathbold}{bold}{U}{BOONDOX-ds}{b}{n}
\DeclareMathAlphabet{\mathcalboondox}{U}{BOONDOX-calo}{m}{n}
\SetMathAlphabet{\mathcalboondox}{bold}{U}{BOONDOX-calo}{b}{n}
\DeclareMathAlphabet{\mathbcalboondox}{U}{BOONDOX-calo}{b}{n}

\DeclareFontFamily{U}{matha}{\hyphenchar\font45}
\DeclareFontShape{U}{matha}{m}{n}{ <5> <6> <7> <8> <9> <10> gen * matha
                    <10.95> matha10 <12> <14.4> <17.28> <20.74> <24.88> matha12}{}
\DeclareSymbolFont{matha}{U}{matha}{m}{n}
\DeclareMathSymbol{\varleftarrow}{3}{matha}{"D0}
\DeclareMathSymbol{\varrightarrow}{3}{matha}{"D1}
\DeclareMathSymbol{\simeq}{3}{matha}{"14}
\DeclareMathSymbol{\sim}{3}{matha}{"12}
\DeclareMathSymbol{\ll}{3}{matha}{"21}
\DeclareMathSymbol{\gtrsim}{3}{matha}{"C1}
\DeclareMathSymbol{\lesssim}{3}{matha}{"C0}

\newcommand\linkcol{RRed}



\usepackage[breaklinks=true,backref=page]{hyperref}
\hypersetup{
    bookmarks=true,         
    bookmarksnumbered=true,
    pdftoolbar=true,        
    pdfmenubar=true,        
    pdffitwindow=false,     
    pdfauthor={},     
    pdfsubject={},   
    pdfcreator={},   
    pdfproducer={},  
    pdfpagemode={UseNone},
    pdfstartview={FitH},
    colorlinks=true,
    plainpages,
    linktoc=page,
    citecolor=blue,
    filecolor=black,
    linkcolor=\linkcol,
    urlcolor=Green,
}
\renewcommand*{\backref}[1]{}
\renewcommand*{\backrefalt}[4]{%
\ifcase #1 %
\relax
\or
~{\small [\textsc{p.~\fns{\!#2}}]}
\else
~{\small [\textsc{p.~\fns{\!#2}}]}%
\fi}

\usepackage{footnotebackref}

\usepackage{cleveref} 

\makeatletter
\normalsize
\makeatother

\def\+{~+~}
\def\-{~-~}
\def\={\:=\:}
\newcommand\fns{\footnotesize}

\newcommand\qqquad{\quad\quad\quad}

\newcommand\Real{\textrm{Re}}

\newcommand\Id{\mathds{1}}

\newcommand\nm{\mathrm{nm}}

\newcommand\Kelv{\mathrm{K}}

\newcommand\eV{\textrm{eV}}

\newcommand\vF{v_\textsc{f}}

\newcommand\Cl{\mathcal{C}_\mathsmaller{\lambda}}
\newcommand\Clp{\mathcal{C}_\mathsmaller{\lambda'}}
\newcommand\FFD{\mathcal{F}_{{\!}_{\textsc{FD}}}}

\newcommand\muchem{\bar{\mu}_{{}_{\!\textsc{c}}}}
\newcommand\Gphi{\mathscr{G}_{\varphi}}
\newcommand\Gz{\mathscr{G}_z}
\newcommand\WW{\mathcal{W}(\Omega)}
\newcommand\zetaN{\zeta_{{}_\ms{N}}}
\newcommand\IeS{\mathcal{I}_\ms{S}}
\newcommand\kdot{k_{\ml{\centerdot}}}
\newcommand\kdotzero{k_{\ml{\centerdot}}^{0}}

\newcommand{\Resintra}{\Real[\sigma^{\text{intra}}]}
\newcommand{\Respp}{\Real[\sigma_{\!{\varphi\varphi}}]}
\newcommand{\Reszz}{\Real[\sigma_{\!zz}]}
\newcommand{\Resinterpp}{\Real[\sigma^{\text{inter}}_{\!{\varphi\varphi}}]}
\newcommand{\Resintrapp}{\Real[\sigma^{\text{intra}}_{\!{\varphi\varphi}}]}
\newcommand{\Resinterzz}{\Real[\sigma^{\text{inter}}_{\!zz}]}
\newcommand{\Resintrazz}{\Real[\sigma^{\text{intra}}_{\!zz}]}

\newcommand{\ms}{\mathsmaller}

\newcommand{\ml}{\mathlarger}

\newcommandx{\dd}[1][1=\mu,usedefault]{\partial_{#1}}
\newcommandx{\ddu}[1][1=\mu,usedefault]{\partial^{#1}}
\newcommandx{\tcR}[1]{\textcolor{Crimson}{#1}}
\newcommandx{\tts}[1]{\text{\textsmaller{#1}}}
\newcommandx{\mlt}[1]{\mathlarger{\text{#1}}}
\newcommandx{\dt}[1][1=f,usedefault]{\frac{\partial{#1}}{\partial t}}
\newcommandx{\dx}[1][1=f,usedefault]{\frac{\partial{#1}}{\partial x}}
\newcommandx{\ddx}[1][1=f,usedefault]{\frac{\partial^2{#1}}{{\partial x}^2}}
\newcommandx{\dm}[1][1=\mu,usedefault]{\partial_{#1}}
\newcommandx{\dmup}[1][1=\mu,usedefault]{\partial^{#1}}
\newcommandx{\subm}[2][1=p,2=A,usedefault]{{#1}_{\!\mathsmaller{#2}}}
\newcommandx{\subt}[2][1=p,2=A,usedefault]{{#1}_\text{\textsmaller{#2}}}
\newcommandx{\supm}[2][1=p,2=A,usedefault]{{#1}^{\!\mathsmaller{#2}}}
\newcommandx{\supt}[2][1=p,2=A,usedefault]{{#1}^\text{\textsmaller{#2}}}
\newcommandx{\subpt}[3][1=p,2=A,3=B,usedefault]{{#1}^\text{\textsmaller{#3}}_\text{\textsmaller{#2}}}
\newcommandx{\subpm}[3][1=p,2=A,3=B,usedefault]{{#1}^{\mathsmaller{#3}}_{\mathsmaller{#2}}}
\newcommandx{\sh}[1][1=\alpha,usedefault]{\sinh\left(#1\right)}
\newcommandx{\ch}[1][1=\alpha,usedefault]{\cosh\left(#1\right)}
\newcommandx{\sech}[1][1=\alpha,usedefault]{\mathrm{sech}\left(#1\right)}
\newcommandx{\cosech}[1][1=\alpha,usedefault]{\mathrm{cosech}\left(#1\right)} \newcommandx{\LCTd}[4][1=\mu,2=\nu,3=\rho,4=\sigma,usedefault]{\eps_{#1#2#3#4}}
\newcommandx{\LCTu}[4][1=\mu,2=\nu,3=\rho,4=\sigma,usedefault]{\eps^{#1#2#3#4}}

\newcommandx{\gmat}[2][1=\mu,2=\nu,usedefault]{g_{{#1}{#2}}}
\newcommandx{\gmatinv}[2][1=\mu,2=\nu,usedefault]{g^{{#1}{#2}}}
\newcommandx{\spc}[3][1=\mu,2=a,3=b,usedefault]{{\omega_{#1}}^{\!\!{#2}{#3}}}
\newcommandx{\Conn}[3][1=\mu,2=\nu,3=\lambda,usedefault]{{\Gamma_{{#1}{#2}}}^{\!\!#3}}
\newcommandx{\viel}[2][1=\mu,2=a,usedefault]{{e_{#1}}^{\!#2}}
\newcommandx{\vielinv}[2][1=a,2=\mu,usedefault]{{e_{#1}}^{#2}}
\newcommandx{\vieluu}[2][1=\mu,2=a,usedefault]{e^{#1#2}}
\newcommandx{\Rdduu}[4][1=\mu,2=\nu,3=a,4=b,usedefault]{{R_{{#1}{#2}}}^{{#3}{#4}}}
\newcommandx{\DD}[1][1=\mu,usedefault]{\mathcal{D}_{#1}}
\newcommandx{\gammafl}[1][1=\gamma,usedefault]{\mkern2mu\check{\mkern-2mu#1}}

\newcommandx{\overbar}[1]{\mkern
1.5mu\overline{\mkern-2.0mu#1\mkern-2.0mu}\mkern 1.5mu}
\newcommandx{\overbarcal}[1]{\mkern                   6.0mu\overline{\mkern-5.5mu#1\mkern-1.0mu}\mkern 1.5mu} 